\documentclass[preprint,superscriptaddress]{revtex4-2}
\usepackage{graphicx}
\usepackage{amsmath, amssymb, soul}
\usepackage[usenames]{color}
\usepackage{stackrel}

\begin{document}


\title{Sink vs. tilt penetration into shaken dry granular matter: \\
the role of foundation}

\author{L. Alonso-Llanes}
\email[]{lalonso@fisica.uh.cu}
\affiliation{Group of Complex Systems and Statistical Physics, Physics Faculty, University of Havana, 10400 Havana, Cuba}
\affiliation{Universit\'{e} de Strasbourg, CNRS, Institut Terre et Environnement de Strasbourg, UMR7063, 67000 Strasbourg, France}

\author{G. S\'{a}nchez-Colina}
\affiliation{Group of Complex Systems and Statistical Physics, Physics Faculty, University of Havana, 10400 Havana, Cuba}

\author{A. J. Batista-Leyva}
\affiliation{Group of Complex Systems and Statistical Physics, Physics Faculty, University of Havana, 10400 Havana, Cuba}
\affiliation{Instituto Superior de Tecnolog\'ias y Ciencias Aplicadas (InSTEC). 
University of Havana, 10400 Havana, Cuba}

\author{C. Cl\'{e}ment}
\affiliation{Universit\'{e} de Strasbourg, CNRS, Institut Terre et Environnement de Strasbourg, UMR7063, 67000 Strasbourg, France}

\author{E. Altshuler}
\affiliation{Group of Complex Systems and Statistical Physics, Physics Faculty, University of Havana, 10400 Havana, Cuba}

\author{R. Toussaint}
\email[]{renaud.toussaint@unistra.fr}
\affiliation{Universit\'{e} de Strasbourg, CNRS, Institut Terre et Environnement de Strasbourg, UMR7063, 67000 Strasbourg, France}
\affiliation{SFF PoreLab, The Njord Centre, Department of Physics, University of Oslo, P.O. Box 1074 Blindern, 0316 Oslo, Norway}


\date{\today}

\begin{abstract}
We study the behavior of cylindrical objects as they sink into a dry
granular bed fluidized due to lateral oscillations. Somewhat unexpectedly, 
we have found that, within a large range of lateral shaking powers,
cylinders with flat bottoms sink vertically, while those with a
\textquotedblleft foundation\textquotedblright consisting in a
shallow ring attached to their bottom, tilt besides sinking. The
latter scenario seems to dominate independently from the nature of
the foundation when strong enough lateral vibrations are applied. We
are able to explain the observed behavior by quasi-2D numerical
simulations, which also demonstrate the influence of the intruder's aspect ratio. The vertical sink dynamics is explained with the help of a Newtonian equation of motion for the intruder. Our findings may shed light on the behavior of buildings and other man-made constructions during  earthquakes.
\end{abstract}


\maketitle

\section{Introduction}

The Kocalei earthquake occurring on August 17, 1999 affected in
various ways many constructions in the city of Adapazari, Turkey.
Following observers, some buildings sank vertically into the soil,
others tilted, and some even suffered lateral translation over the ground 
\cite{Bray1999,Sancio2002,Sancio2004}. This case illustrates well 
the diversity of damage that earthquake fluidization of soils may 
cause to man-made structures \cite{Ambraseys1988}. 

Liquefaction in the ground may be triggered dynamically by
waves emitted during earthquakes, generating cyclic shear stresses 
that lead to the gradual build-up of pore water pressure \cite{BenZeev2017, BenZeev2020}. 
The shaking produced by seismic events is a trigger for extensive liquefaction, 
as was observed recently in Belgium \cite{Vanneste1999}.
Ground fluidization \cite{NRC1985,Berrill1985} has been investigated
in different kinds of media like sand \cite{Berrill1985}, dry
granular soils \cite{Clement1991} and sediments \cite{Wang2009}. Of
immediate interest for engineering and for the geosciences is to
understand how man-made structures such as buildings, and massive
rocks laying on granular soils respond to fluidization associated to
seismic waves.

\begin{figure}
\includegraphics[width=0.4\textwidth]{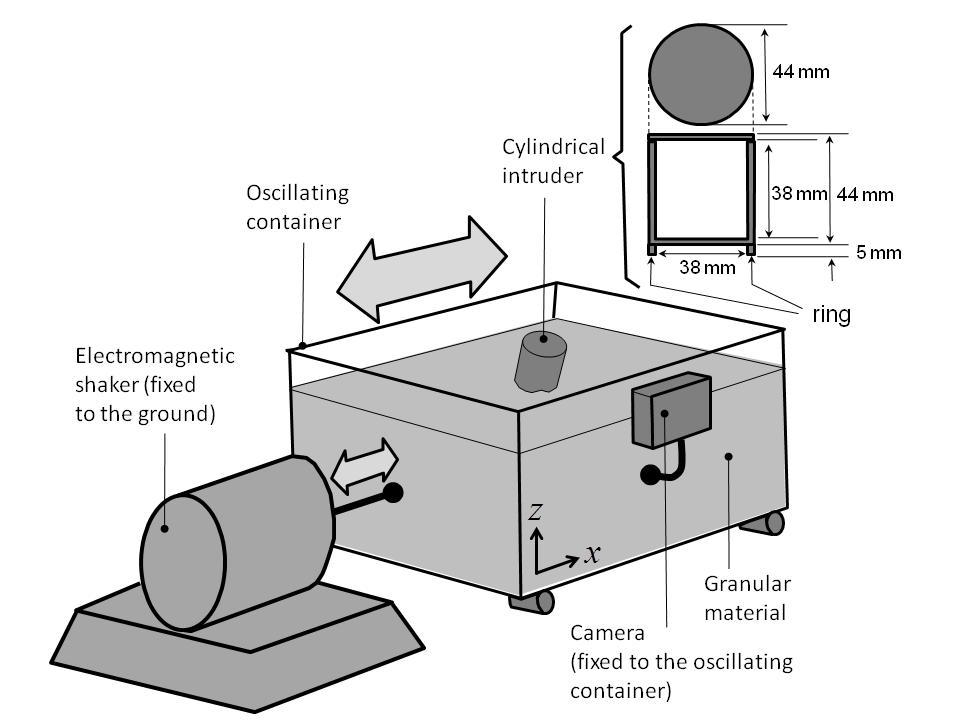}
\caption{\label{Fig1} Experimental setup. At the upper right, we have
illustrated the intruder consisting in a cylinder with ring.}
\end{figure}

Granular matter itself displays a variety of puzzling phenomena
\cite{Jaeger1996,Shinbrot1998,Altshuler2003PRL,Aranson2006,Andreotti2013,Altshuler2008PRE,Boudet2007,Vinningland2007,Johnsen2006,Johnsen2008,Niebling2010a,Niebling2010b,Niebling2012,Turkaya2015}, 
but during 
the last decade or so, our understanding of the dynamics of objects 
penetrating into granular media has advanced quickly 
\cite{Uehara2003,
Boudet2006,Katsuragi2007,Goldman2008,Pacheco2011,
Kondic2012,Torres2012,Clark2013,Ruiz2013,Brzinski2013,Altshuler2014,Joubaud2014,Harich2011, Kolb2004,Li2013,Aguilar2016,Kang2018,Diaz2020}.
While laterally shaken granular beds have received a certain degree of
attention \cite{Medved2000,Kruelle2009}, the performance of
objects initially laying on the surface of a granular bed submitted
to lateral shaking has been rarely studied
\cite{Tennakoon1999,Liu1997,Sanchez2014,Clement2018p,Clement2018}.

In this paper we perform systematic experiments associated to the
latter scenario, which may help understanding the performance of
human constructions and rocks  laying on granular beds during
earthquakes. In particular, using a cylinder as a simplified model,
we study its settling dynamics on a granular
bed submitted to lateral vibrations. Somewhat unexpectedly, we have
found that, within a large range of lateral shaking powers,
cylinders with flat bottoms sink vertically, while those with a
\textquotedblleft foundation\textquotedblright  consisting in a
shallow ring attached to their bottom, tilt besides sinking. The
latter scenario seems to dominate independently from the nature of
the foundation when strong enough lateral vibrations are applied.
Quasi-2D simulations mimicking the experiments were also performed.
The settling dynamics of the simulated intruders, with or without
foundation, reproduces the corresponding experimental
results. Our simulations also reveal how the difference in force-chain 
distributions between flat and non-flat bottom cylinders produces different torques 
justifying the two types of penetration. In addition, we present a simple 
phenomenological model that reproduces well the sinking
dynamics and helps understanding how the tilting process influences 
the sinking one.

\section{Experimental}

The penetration experiments were performed on a granular bed
contained in a test cell of approximately $25 \times 25 \times
25$~cm$^3$ filled with Ugelstad spheres of non expanded polystyrene
with a bulk density 1050 kg/m$^3$, and diameter 140 $\mu$m (monodisperse
within a $1$ percent), type Dynoseeds \textcopyright , produced by
Microbeads, Norway. The box was horizontally shaken at different
amplitudes of motion ($A$), and a frequency ($f$) of  $5.0$ Hz (a value
commonly found in seismic waves), using a TIRA \textcopyright TV51120-M shaker, see figure \ref{Fig1}. By controlling the voltage of the shaker input signal we varied the amplitude of the oscillations up to a maximum value corresponding to a peak ground acceleration of $A(2 \pi f)^2 \approx 12.2$ m/s$^2$ {\cite{limit}}. This acceleration range covers most potentially damaging earthquakes, from weak to strong \cite{USGS}, though there has been reports of larger peak ground accelerations {\cite{Goto2019}}. The time the shaker needs to reach the steady state depends on the dimensionless acceleration, being longer for larger accelerations. The time intervals can range from one (0.2 s) to three periods.

Two types of intruders were used in the experiments: (a) a hollow 3D printed
cylinder of $44$~mm diameter, $44$~mm height $h_c$ (external dimensions), and $5$~mm
thick walls, and (b) the same cylinder with a ring of $5$~mm height and
$3$~mm thickness glued to its bottom (illustrated in the upper right
corner of Fig. \ref{Fig1}). Intruders (a) and (b) will be called
\textquotedblleft No-ring\textquotedblright and \textquotedblleft Ring\textquotedblright,
respectively, from now on. Their masses were adjusted with ballast in such a
way that their densities matched the average effective density of the granular
medium, which was measured as $430$ kg/m$^3$. As far as the ballast used has a
density near the effective density of the granular material, it was
almost evenly distributed inside the cylinder. Note that, using a flat
bottom cylinder and a ring-like bottom cylinder, we are modifying the
\textquotedblleft foundation\textquotedblright of our intruder.

A digital camera \textit{Hero 2} made by GoPro \textcopyright \ was fixed to the
electromagnetic shaker, in such a way that it could take a video of
the sinking process from an oscillating reference frame locked to
the test cell, as proposed in \cite{Sanchez2014}. This method
allowed a much better quality of the cylinder's images, and made
easier their digital processing. Videos were taken at a maximum rate
of $120$ frames per second, with a resolution of $1920 \times 1080$
pixels.

\begin{figure}
\includegraphics[width=0.45\textwidth]{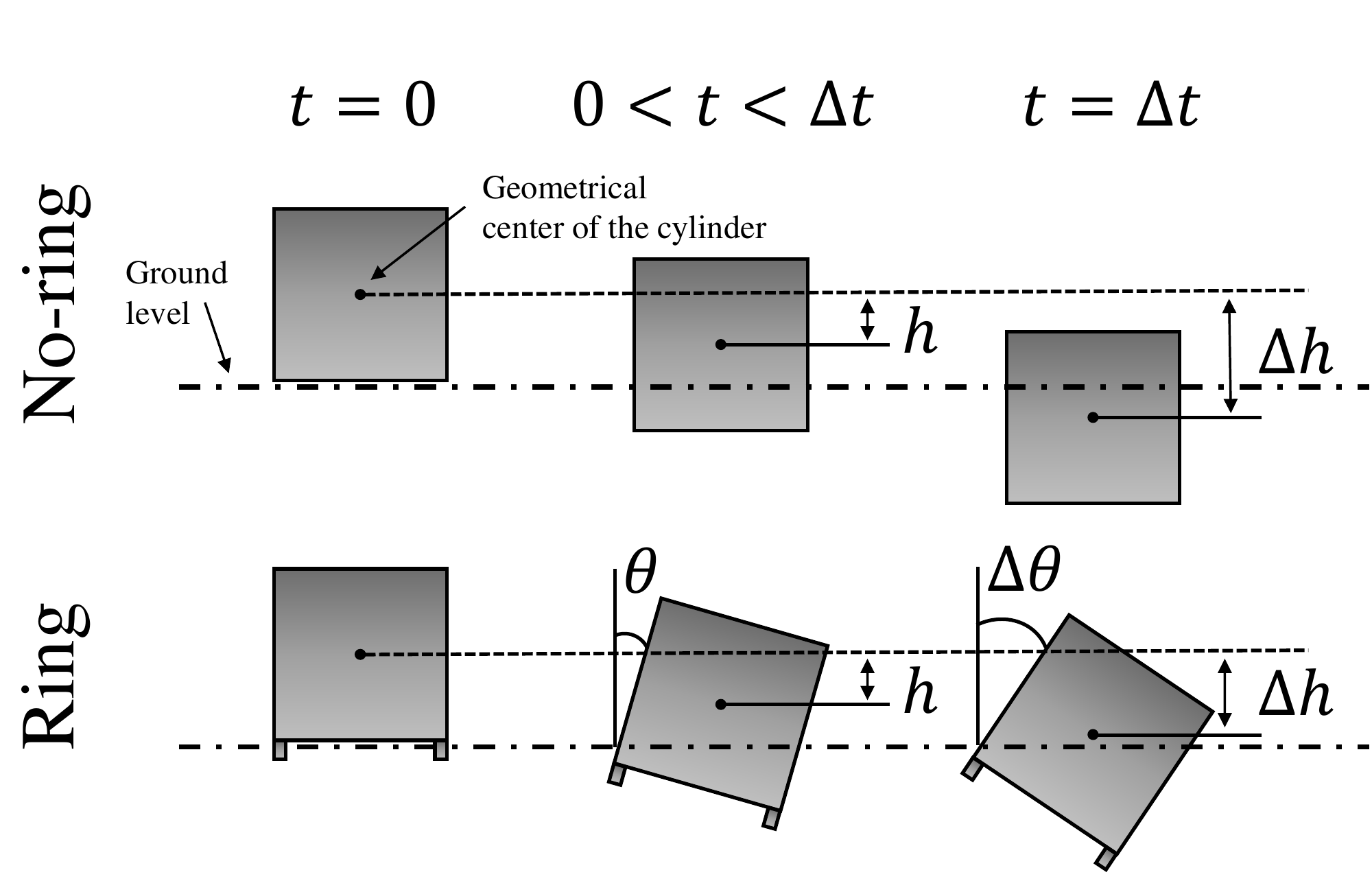}
\centering \caption{\label{Fig2} Sketch of sinking and tilting processes. The top 
row illustrates the sinking process of a No-ring cylinder in three 
moments during the experimental interval from $t=0$ to a final time  
$t=\Delta t$. The bottom row shows the same temporal sequence for a 
Ring cylinder, which tilts in addition to sink.}
\end{figure}

The images were processed as follows. We first converted the videos to
image sequences in *.jpg format, and cropped each picture, excluding
irrelevant space. Then, the images were binarized through an appropriate
threshold. Using the tool \textit{regionprops} from \textit{MatlabR2014a} 
\textcopyright , we
identified and assigned coordinates to several bright marks we had glued
to certain points of the cylindrical intruder. The coordinates of the
marks were used to calculate the position of the intruder's geometrical
center and inclination relative to the vertical in each picture. In some
experiments where the sinking was particularly big, it was difficult to
obtain the tilt angle, since part of the marks sank below the level of the
sand surface, and they were impossible to follow. In such cases the upper
border of the cylinder was identified using the Matlab's tools \textit{find} and
\textit{bwtraceboundary}, and then fitted to a polynomial using the function
\textit{polyfit}. The fit was used to find the inclination. In the case of
experiments ending in a very inclined position, the reference to calculate
the inclination was the cylinder's corner above the sand surface, that was
identified as the intersection of the two polynomial fits of the upper and
one lateral borders of the cylinder.

As the cylinder oscillates due to the vibration of the box, it is difficult
to determine the final position, particularly when there is a big tilting.
Then, in order to determine the sinking depth and tilting, we observe in the
videos the onset of a cyclic  movement of a reference point in the cylinder. 
Then, the final position could be measured in the frames filmed after the 
shaker was stopped.

The experimental protocol  can be described as follows: (I) preparing
the granular medium by stirring it evenly with a long rod, (II)
gently depositing the cylinder in the upright position on the free
surface of the granular bed, (III) turning ON the camera, (IV)
switching ON the shaker after setting the desired frequency and
amplitude (V) turning OFF the shaker and the camera after the
penetration process had visibly finished.

In Fig. \ref{Fig2} we define the main parameters describing the sinking process of
a No-ring cylinder (upper row), and the tilting and sinking of a Ring
cylinder (bottom row), during the experimental lapse, defined as $\Delta t$. As
the figure indicates, in the following we will call $h$ the penetration of the
geometrical center at a time $t$ and $\Delta h$ the final penetration at time $t=\Delta t$.
Note that both magnitudes are defined as the vertical displacement of the geometrical
center of the cylinder (without taking the ring into account).
In the same way we will call $\theta$ the inclination of the intruder at time $t$ and 
$\Delta \theta$ the final inclination at $t = \Delta t$.

\begin{figure}
\includegraphics[width=0.45\textwidth]{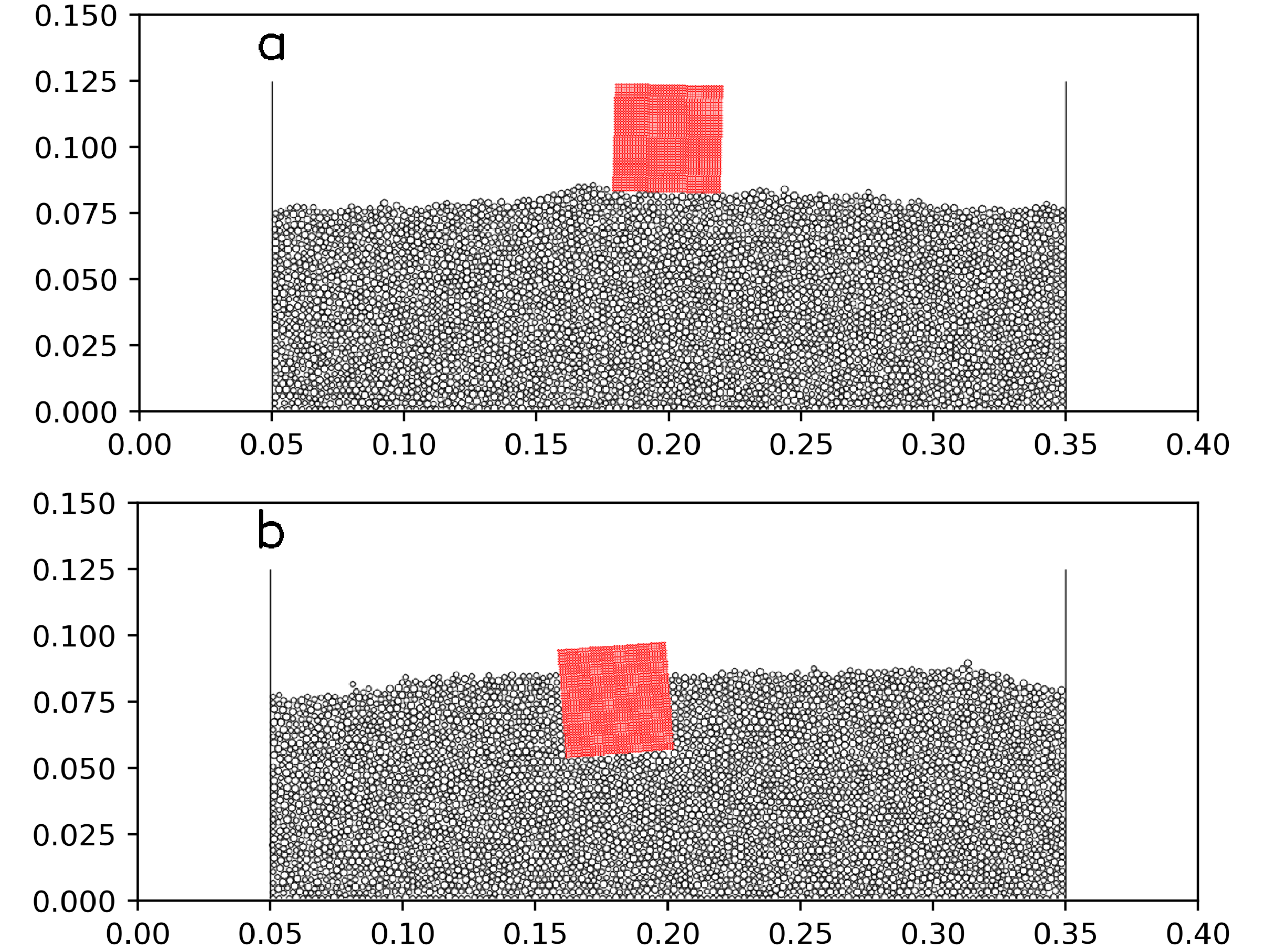}
\centering\caption{\label{Fig3}(color online) Snapshots of the initial (a) and final (b)
positions of a No-ring intruder in a typical quasi-2D simulation using
a shaking frequency of 5 Hz.}
\end{figure}
\begin{figure}
\includegraphics[width=0.45\textwidth]{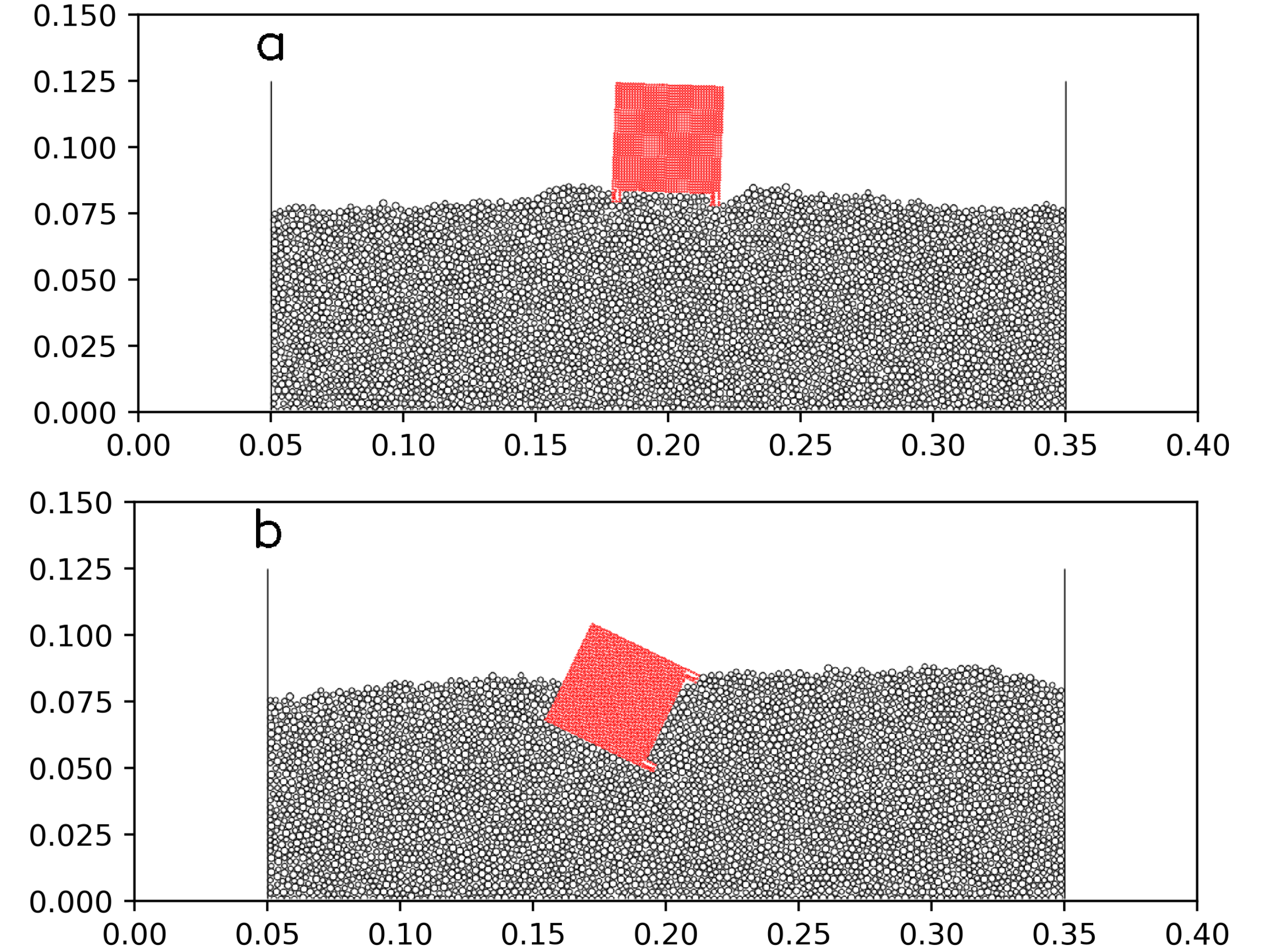}
\centering \caption{\label{Fig4}(color online) Snapshots of the initial (a) and final (b)
positions of a Ring intruder in a typical quasi-2D simulation using a
shaking frequency of 5 Hz. }
\end{figure}

We also explored the phenomenology through numerical
simulations. They were based on a discrete element method code (DEM)
for the computation of granular systems \cite{Cundall1979,Parez2016,Johnsen2006,
Niebling2010a,Niebling2012,Clement2018p,Clement2018}.
We modeled a quasi-2D granular medium, made of finite-sized hard spheres 
with radii between 1.0 and 1.5 mm, to avoid the crystallization effect. 
The medium contains 4000 particles and is prepared by placing the 
latter randomly in a space 30 cm wide and 25 cm high and then allowing 
them to settle under the action of gravity $g=9.81$ m/s$^2$. Once the 
medium reaches equilibrium, it occupies a virtual space 30 cm wide and 
about 8 cm high, laterally delimited by flat walls that define the Hele-Shaw cell.
The simulation box used was created narrow in order to have a single 
plane of particles in the direction perpendicular to the images shown 
in Figures \ref{Fig3} and \ref{Fig4}. The components of the velocities 
and forces along this direction are set to zero at each time step.
To mimic the experimental conditions, we simulate particles of density 1050 kg/m$^3$.

The two intruders are made of cohesive particles. One is a square
of 40 mm side, made of $N = 1681$ particles with diameter 1 mm, placed in a 
quasi-2D square arrangement, which simulates the No-ring intruder of the
experiments. The second one is also a square of 40 mm side to which two
\textacutedbl small feet\textacutedbl are attached. Each foot has a size 
of 4.5 $\times$ 2.7 mm$^2$, so the simulated Ring intruder contains a 
total of $N = 1705$ particles. 
The density of the spheres $\rho_p$ which form the intruders is adjusted 
so that the bulk density of the rigid body matches the effective density of the granular medium.
The latter is calculated once the medium has settled down and is stable,
and was always found to be around $\rho_m = 566$ kg/m$^3$. 
Then, the density of the particles forming the intruder is obtained
as $\rho_p = \rho_m V_i / N V_p$, where $V_i$ is the volume of the intruder
and $V_p$ the volume of a sphere.

Once our granular medium is created, we place the intruder 1 mm
above the medium, with its bottom parallel to the horizontal direction.
We release it, under the action of the force of gravity, and wait until the whole
system becomes motionless (i.e. its total kinetic energy reaches a value under $10^{-7}$ J).
Then, we apply horizontal oscillations of different amplitudes and a 
frequency of 5 Hz ($\Delta t\approx 8$s) 
to the walls of the cell and compute the time evolution of the 
position and tilting angle of the intruder. The amplitudes were chosen 
in such a way that the dimensionless acceleration $\Gamma = A (2 \pi f)^2 / g$ 
had the values of 0.16, 0.25, 0.5, 1.0, 1.25 and 1.5. 

The contact between spheres was modeled as a linear spring-dashpot $F_{ij} = (k_n \delta \mathbf{n}_{ij} - m_{eff}\gamma_n\mathbf{v}_n) - (k_t \Delta \mathbf{S}_t + m_{eff} \gamma_t \mathbf{v}_t)$ \cite{Cundall1979, Thompson2021}, where $k$ and $\gamma$ are the elastic and viscoelastic damping constants, $\delta \mathbf{n}_{ij}$ is the overlap distance along the line connecting the centers of the two spheres, and $\mathbf{v}$ their relative velocity. $\Delta \mathbf{S}_t$ is the tangential displacement vector between two spheres, which is truncated to satisfy a frictional yield criterion, and $m_{eff} = m_i m_j / (m_i + m_j)$ is the effective mass of two spheres of mass $m_i$ and $m_j$. We considered normal ($n$) and tangential ($t$) forces components between the particles and,
in order to model hard spheres that interact on contact (i.e. spheres 
whose deformation during collisions is less than 
a small fraction of their radii), we used the following parameter values: 
$k_n = 1.2 \times 10^7$ N/m, $k_t = 2/7 k_n$, $\gamma_n = 12$ s$^{-1}$ and $\gamma_t = 0.1 \gamma_n$.
The interaction force between the walls and the particles touching them is
the same as the corresponding for two particles but considering the wall 
of infinite radius and mass (flat wall). The microscopic friction coefficient 
between spheres, and between spheres and boundaries, was taken as $\mu = 0.3$.
The time step $dt$ was chosen to guarantee that there are at least 50 steps during 
one characteristic time of a collision $dt = t_c/50$, where $t_c = \pi/\sqrt{(k_n/m_{eff}) - \gamma_n^2}$.

Figures \ref{Fig3} and \ref{Fig4} show the initial and final positions 
of both types of intruders in two typical runs. Fig. \ref{Fig3} indicates 
that the No-ring cylinders are slightly inclined, while in Fig. \ref{Fig4}
the large inclination of the Ring one becomes obvious.

\begin{figure}
\includegraphics[width=0.45\textwidth]{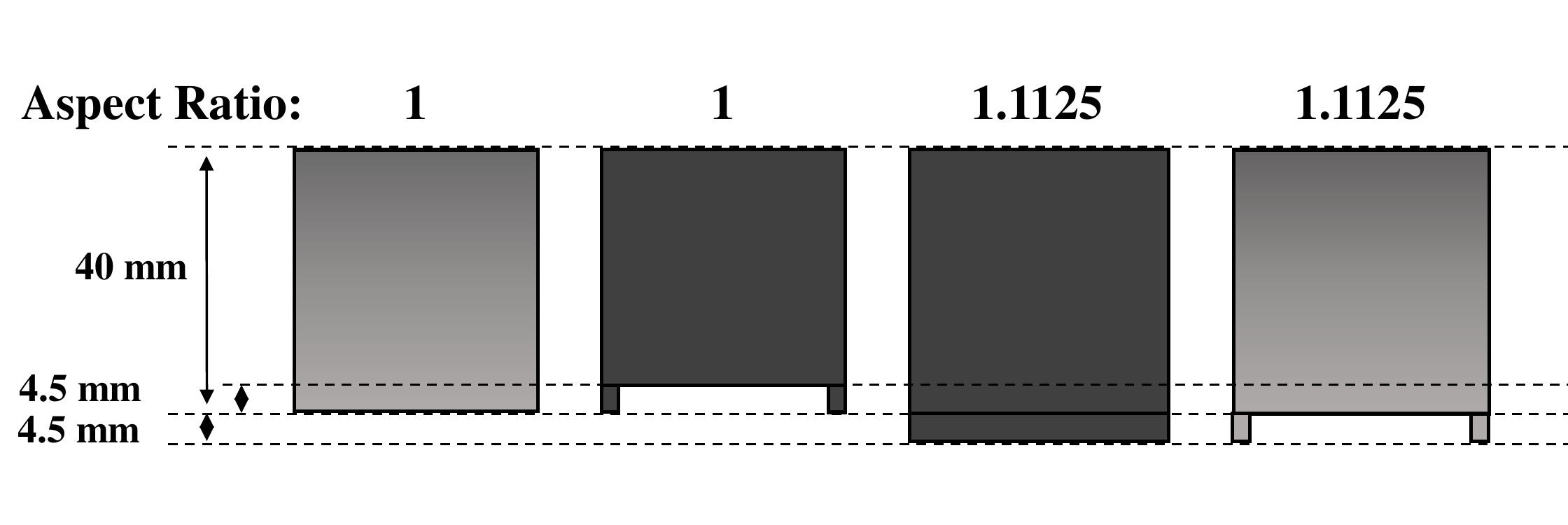}
\centering \caption{\label{Fig4.1} Comparison of the geometry of the intruders used to test the influence of the aspect ratio. In light gray the intruders previously described and in dark gray the new ones.}
\end{figure}

We also performed an additional set of simulations aimed at elucidating the influence of the intruder's aspect ratio in the sink - tilt behaviour. Two intruders, one No-ring with aspect ratio 1.125, and a Ring with aspect ratio 1 (see Fig. {\ref{Fig4.1}}), were submitted to the same range of dimensionless accelerations. Note that the new intruders (in dark gray in Fig. {\ref{Fig4.1}}) have the same dimensions of the former ones (in light gray) but the geometry of the bottom is interchanged.

The new intruders are also rigid bodies made of 1 mm diameter cohesive particles and their bulk densities also correspond to that of the granular medium. The new Ring intruder is composed by $N = 1498$ particles forming a 40 mm wide and 35.5 mm high rectangle to which two $4.5 \times 2.7$ mm$^2$ feet were added. Note that the size of these feet and those of the Ring intruder are the same. The new No-ring is a rectangle of 40 mm wide and 44.5 mm high composed of $N = 1845$ particles.

\section{Results and discussion}

\subsection{Sink \textit{vs.} tilt penetration in experiments}

\begin{figure}
\includegraphics[width=0.45\textwidth]{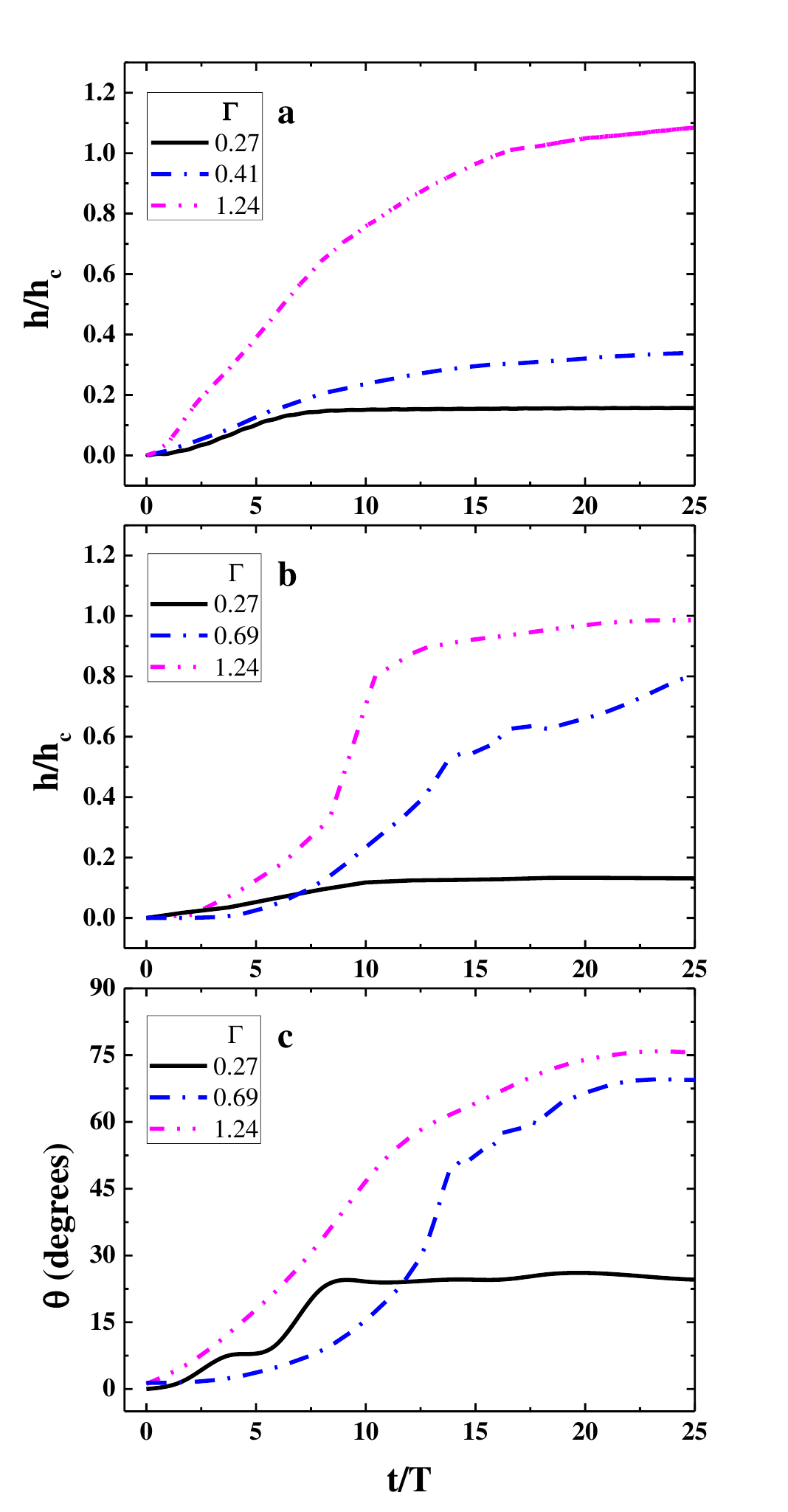}
\centering \caption{\label{Fig5}(color online) Experiments. Time evolution of penetration
depths and tilt angles. Time dependence of the penetration depth of
a No-ring cylinder (a), the penetration depth of a Ring cylinder (b)
and the tilting angle of a Ring cylinder (c), for different
dimensionless accelerations. The long-time creep process is not
completely shown. The tilting angle of No-ring cylinders is not
displayed, due to the fact that it oscillates around angles not
larger than $5\textrm{\textdegree}$ relative to the vertical direction.}
\end{figure}

Figure \ref{Fig5}(a) shows the time variation of the sinking depth
for selected values of the dimensionless acceleration $\Gamma = A (2
\pi f)^2 / g$ (where $g=9.81$ m/s$^2$ is the gravitational
acceleration and $ A (2 \pi f)^2 $ is the horizontal peak
acceleration of the sand box) for No-ring cylinders. It is easy to
see that the penetration of the No-ring cylinders follows a common
pattern for all the accelerations. A first process of fast sinking
is followed by a slow creep. Only the penetration depth increases
with $\Gamma$. In this figure we do not show the total creep
process, due to its long duration. As the height of the cylinder is $h_c = 44$
mm, it is possible to check from Fig. \ref{Fig5}(a) that, for a
dimensionless accelerations of 1.24, the cylinder sinks completely.
An important characteristic of the sinking process in this type of 
cylinder is that the intruder penetrates the granular medium with 
almost no tilting, and a final inclination smaller than 5\textdegree.

Figure \ref{Fig5}(b) is similar to the previous one, but measurements 
were performed with Ring cylinders.
The general features of both graphics are similar, but there is
a difference, that will be better observed in the following figures:
the dimensionless acceleration at which the cylinder sinks completely
in the medium is bigger for the Ring cylinders than for the No-ring
ones. 

Figure \ref{Fig5}(c) presents the time evolution of the tilting angle
for a Ring cylinder, a process that occurs simultaneously with the 
sinking. The sinking and tilting
dynamics of Ring cylinders is more irregular than that of the
No-ring ones. This is illustrated in Fig. \ref{Fig5}(b) and (c),
even after being submitted to a moving average process, to get a smoother 
graph.  

\begin{figure}
\includegraphics[width=0.45\textwidth]{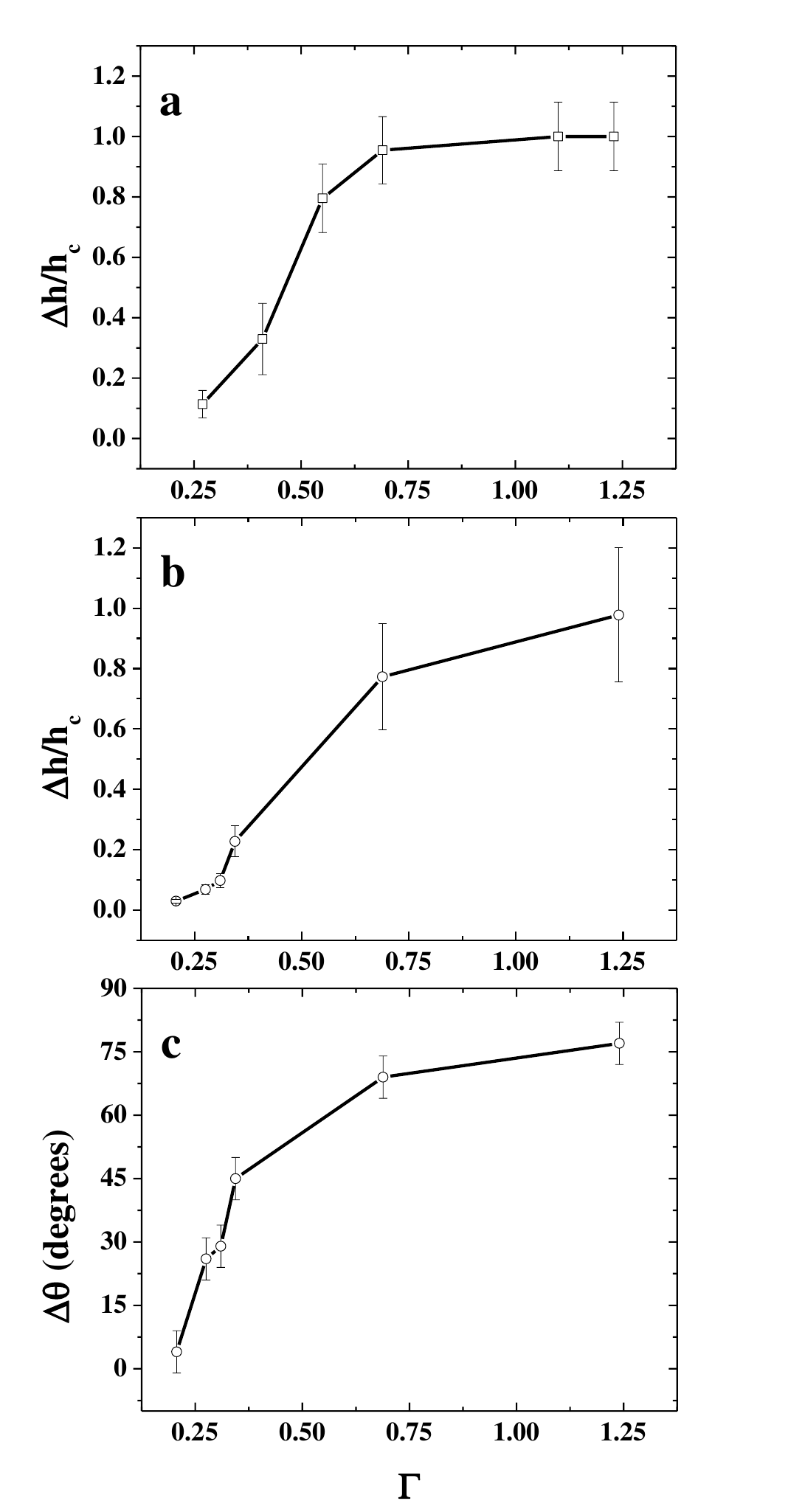}
\centering \caption{\label{Fig6} Experiments. Sinking and tilting: heights and angles for different
dimensionless accelerations $\Gamma$. Final sink heights for No-ring cylinders (a) and for Ring
cylinders (b). Final tilt angles for Ring cylinders (c). Tilt angles of
No-ring cylinders are not shown for the same reasons of the previous
figure.} 
\end{figure}

No-ring cylinders tend to sink vertically as the granular soil is
fluidized by horizontal shaking, while cylinders with rings tend to tilt. Figure
\ref{Fig6} quantifies the differences between the initial and final stages of the
process, for almost all the range of accelerations our experimental setup was able to reach.

Figure \ref{Fig6}(a) shows sink data for No-ring cylinders. As can
be seen, for dimensionless accelerations up to $\Gamma$ = 0.27, there
was no significant penetration of the intruder into the granular
bed. Vertical penetrations started to increase significantly above
this acceleration, reaching a plateau around $\Gamma \approx
0.7$. At the plateau, the cylinder has sunk completely, but stays
\textquotedblleft floating\textquotedblright \hspace{1 mm} into the fluidized
granular medium, as expected for an object isodense relative to it, so 
there is no further sinking. 

In Fig. \ref{Fig6}(b) the sinking process of the Ring cylinders is
summarized. Though the low acceleration part is similar to Fig. \ref{Fig6}(a),
now the plateau is not
observed for the range of accelerations recorded. Notice that from the depth reached at $\Gamma \approx$ 1.2, approximately the height of the cylinder $h_c$, it would not sink any further, and that from this value of $\Gamma$ onwards a plateau would appear.

Figure \ref{Fig6}(c) shows the tilt data for Ring cylinders.  No
significant tilting is observed for $\Gamma$ smaller than
approximately 0.25. With the increase of the dimensionless
accelerations, the cylinder significantly tilts, increasing abruptly
the tilting angle with $\Gamma$, until it slows down at $\Gamma
\approx 0.7$. We do not show
the tilting angle of No-ring cylinders, because it is always smaller
than 5\textdegree, with a random distribution of values around the
vertical direction.

Figures \ref{Fig6}(b) and (c) are closely related, because they are
two descriptions of the same process: the motion of Ring cylinders
into the granular medium, that includes both sinking and tilting.
The fact that at the accelerations shown in this figure the plateau
in the sinking depth is not reached while for the tilting angle
at higher values of $\Gamma$ the inclination almost saturates, could be explained 
by the increase of the friction of the
intruder with the granular medium when the tilting angle increases.
At $\Gamma \approx 0.7$ the intruder has reached a large inclination, 
but is not completely submerged into
the medium. An increase in the acceleration does not increase
significantly the angle, because the resulting torque has diminished
due to the influence of both sinking and tilting, but the increase
in fluidization helps further sinking, until most of the cylinder is
submerged.
    
The sinking process can be understood taking the experimental results 
in Ref. \cite{Tennakoon1999} into account. When the system is submitted to 
lateral shaking, a fluidized layer appears in the upper part of the granular 
cell. This layer reaches a depth $h_f$  that depends on the dimensionless 
acceleration $\Gamma$. Below this layer exists a
\textquotedblleft solid\textquotedblright \ layer. For accelerations in 
the range spanned in our experiments, $h_f$ varies almost linearly with 
$\Gamma$ (see Fig. 3(a) in Ref. \cite{Tennakoon1999}), so we can write

\begin{equation}
    \label{eq.1}
    h_f(\Gamma)= \alpha (\Gamma - \Gamma^*); \Gamma > \Gamma^*
\end{equation}

\noindent where $\Gamma^*$ is the onset of fluidization and $\alpha$ is the slope
of the linear dependence. If $\Gamma \leq \Gamma^*$ the depth of the fluidized
layer is zero.

Then, at  low values of  $\Gamma$ the granular medium is not
fluidized, and the cylinder almost does not sink (merely 5 mm at
$\Gamma=0.27$; see Fig. \ref{Fig6}(a)). For accelerations above the
fluidization threshold, the cylinder sinks until it gets in contact
with the solid layer. The larger is the acceleration, the deeper is
that layer, so the bigger is $\Delta h$. But as soon as the solid
layer appears at a depth larger than the cylinder's height, it does
not sink further: instead, it \textquotedblleft
floats\textquotedblright \ due to isodensity with the sand, so a
plateau is reached.

According to reference \cite{Toussaint2014}, $\Gamma^*$ can be taken
as proportional to the friction coefficient $\mu$ between the
cylinder and the granular medium. In these experiments we can 
approximate $\mu \approx 0.3$, which is the value we use in the 
simulations. The authors also conclude that the final depth of intrusion 
depends on isostasy, and on the severity of shaking. It can be 
entirely determined by isostasy, when the shaking completely unjam 
the medium and suppresses the average friction around the intruder 
\cite{Clement2018}.

To better understand the differences in the dynamics of both types of 
intruders, we performed numerical simulations and their results are 
described below.

\subsection{Sink vs tilt in quasi-2D numerical simulations}

\begin{figure}
\includegraphics[width=0.45\textwidth]{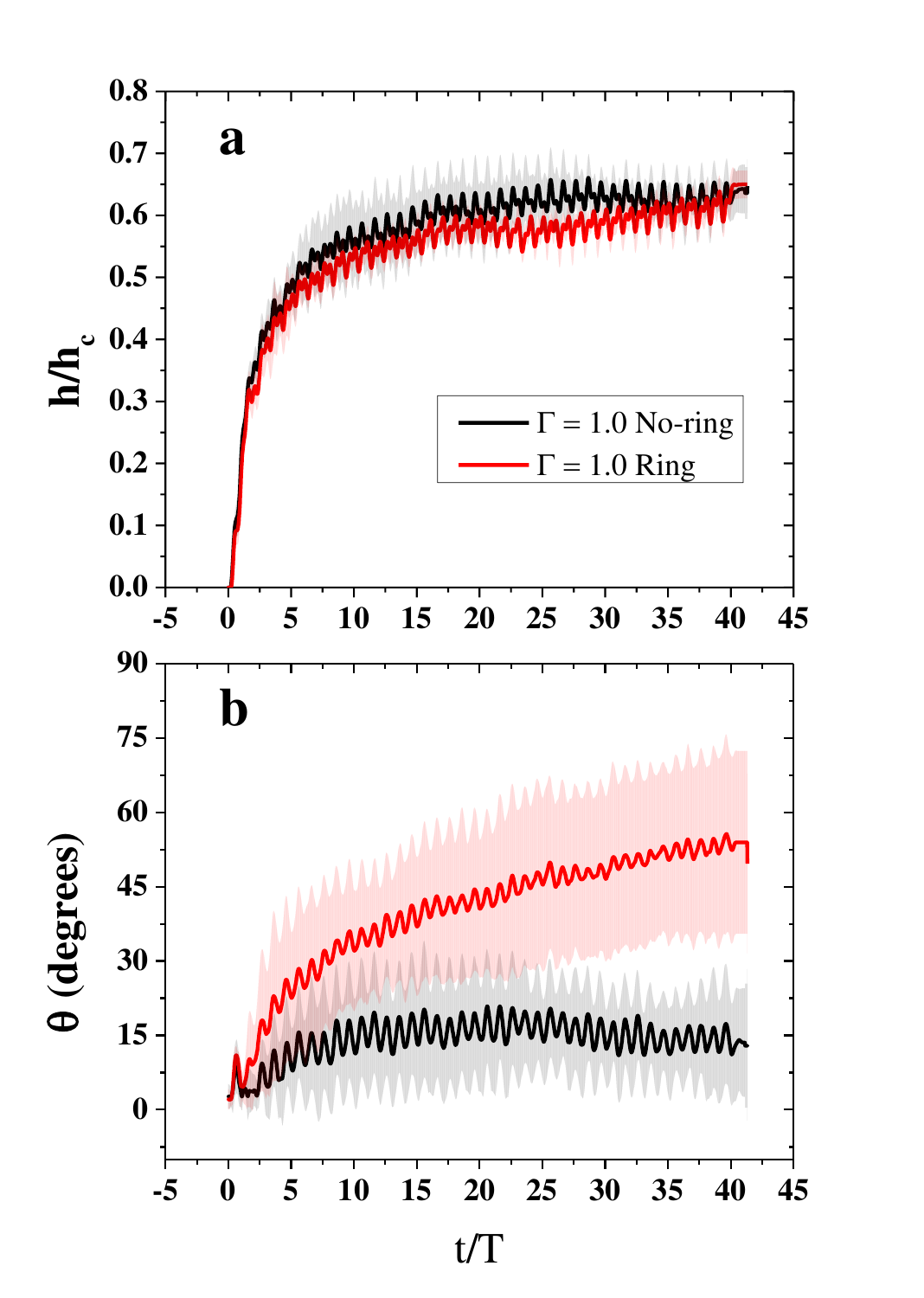}
\centering \caption{\label{Fig7}(color online) Simulations. Time dependence of sinking depth (a)
and tilting angle (b) for both types of intruders with $\Gamma=$ 1.0.
The central lines are the average of six simulations, while the colored
bands cover $\pm$1$\sigma$.}
\end{figure}

Figure \ref{Fig7} shows the time dependence of the penetration depth (a)
and tilting angle (b) for both types of intruders at the dimensionless acceleration 
$\Gamma =$ 1.0. In both figures the thick curves represent the average value of six 
repetitions varying the initial conditions and the surrounding zone represents 
$\pm$1 standard deviation. 

Regarding the vertical sinking in Fig. \ref{Fig7}(a), we do not 
observe major changes between Ring and No-ring intruders; both 
types of intruders sink less than in the experiments. This may be related 
with the lower dimensionality of the 
simulations relative to the real experiment. Quasi-2D granular media allow 
less choices of readjustment than in 3D: they are easily jammed, which 
makes it more difficult for an 
object to sink. Moreover, the size ratio of the intruder over the particles 
is 8 times smaller in the simulations than in the experiments (experiments: 
44 mm/0.140 mm $\approx$ 300; simulations: 40 mm/1 mm= 40), which means that 
if one particle is stuck under the intruder during the simulations, it 
will slow down the intruder significantly more than if the particle were 
8 times smaller. 

Figure \ref{Fig7}(b) indicates that the presence of a 
foundation at the bottom of the intruder causes a large tilting. 
Indeed, for the shaking with no ring, the intruder tilting angle 
is around 10\textdegree, but during the shaking with ring, the intruder 
tilts fast, reaching an angle around 50\textdegree. This resembles what
happens in the experiments (see Fig. \ref{Fig4}): the intruder almost 
ends up lying on one of its sides. Of course, the tilting is also 
limited by the diminished dimensionality in the quasi 2D simulations. 

\begin{figure}
\includegraphics[width=0.45\textwidth]{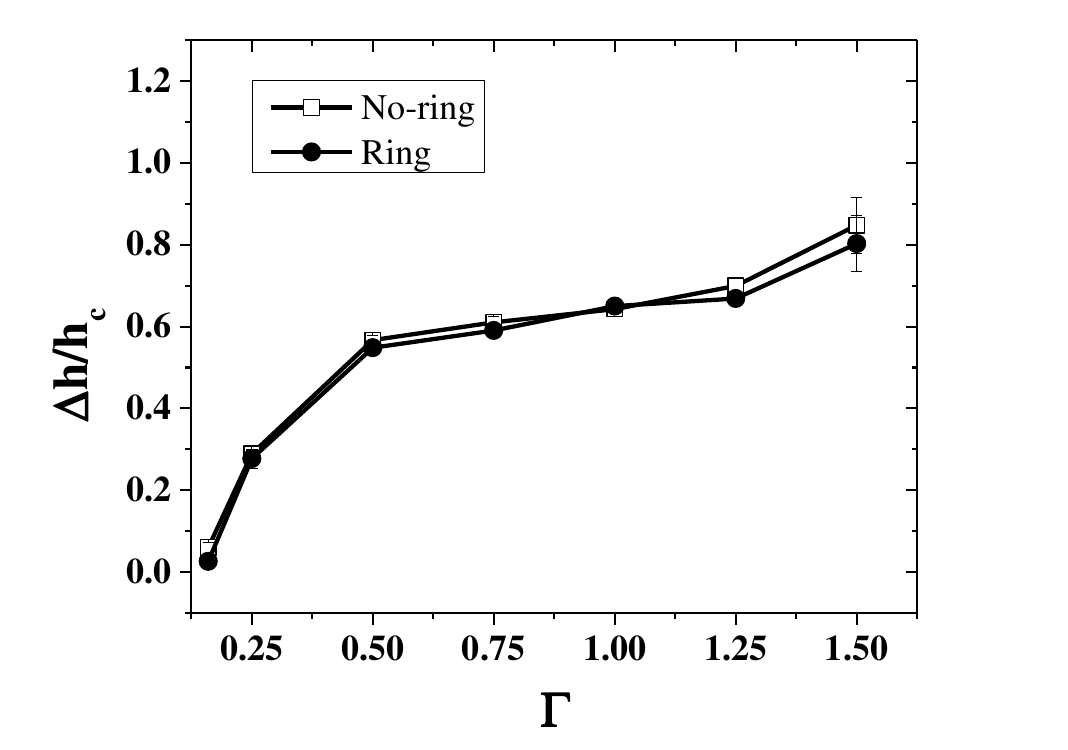}
\centering \caption{\label{Fig8}(color online) Simulations. Final depth reached for No-ring
(squares) and Ring (circles) cylinders as a function of the dimensionless 
acceleration. Symbols represent the average of six experiments and the error
bars $\pm \sigma$.}
\end{figure}

Figure \ref{Fig8} compares the penetration depth reached for both 
types of cylinders at different values of $\Gamma$. The conclusions obtained from
Fig. \ref{Fig7} are valid for all the dimensionless accelerations used in 
the simulations: there are no significant differences in the final sinking depth between
both types of intruders.

\begin{figure}
\includegraphics[width=0.45\textwidth]{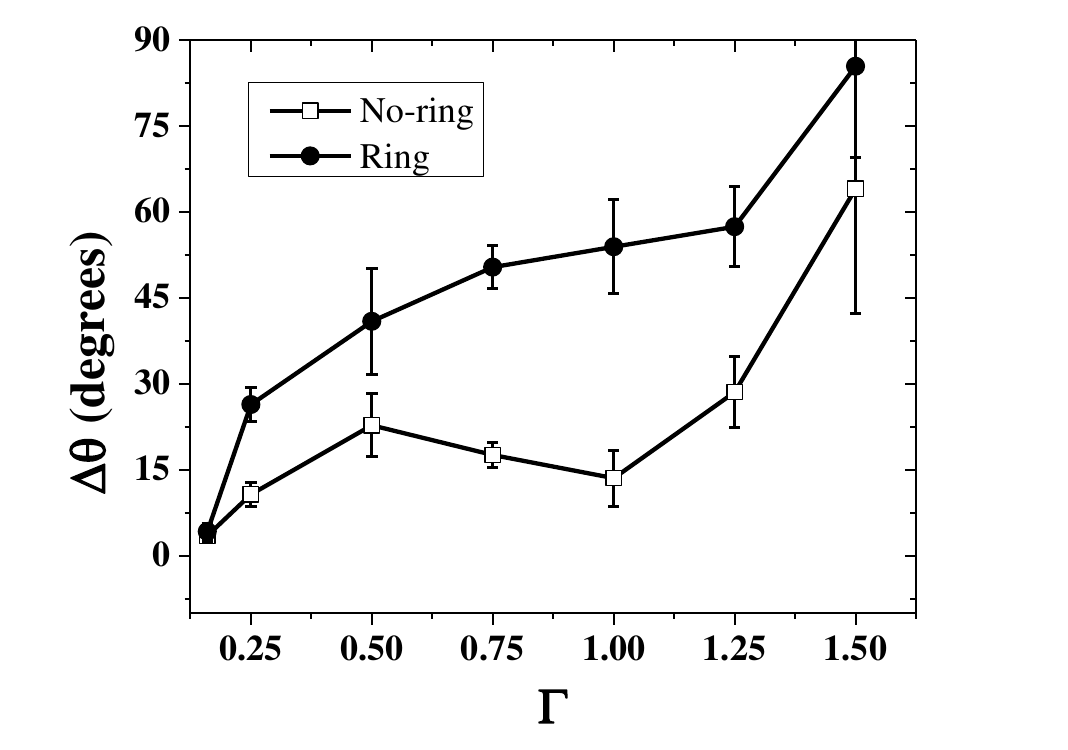}
\centering \caption{\label{Fig9}(color online) Simulations. Maximum tilting angle reached for No-ring
and Ring cylinders as a function of the dimensionless acceleration. Symbols 
represent the average of six simulations and the error bars $\pm \sigma$.}
\end{figure}

In Fig. \ref{Fig9}, on the contrary, the difference in tilting angles between 
the two types of intruders is clearly seen. For all the values of $\Gamma$ the
simulated Ring intruder tilts more than the No-ring one. For $\Gamma$=1.50, the
No-ring tilts up to an angle that is closer to the Ring's one, corresponding to preliminary observations found in experiments with frequencies above 5 Hz for dimensionless accelerations $\Gamma > $1.25.

\begin{figure}
\includegraphics[width=0.4\textwidth]{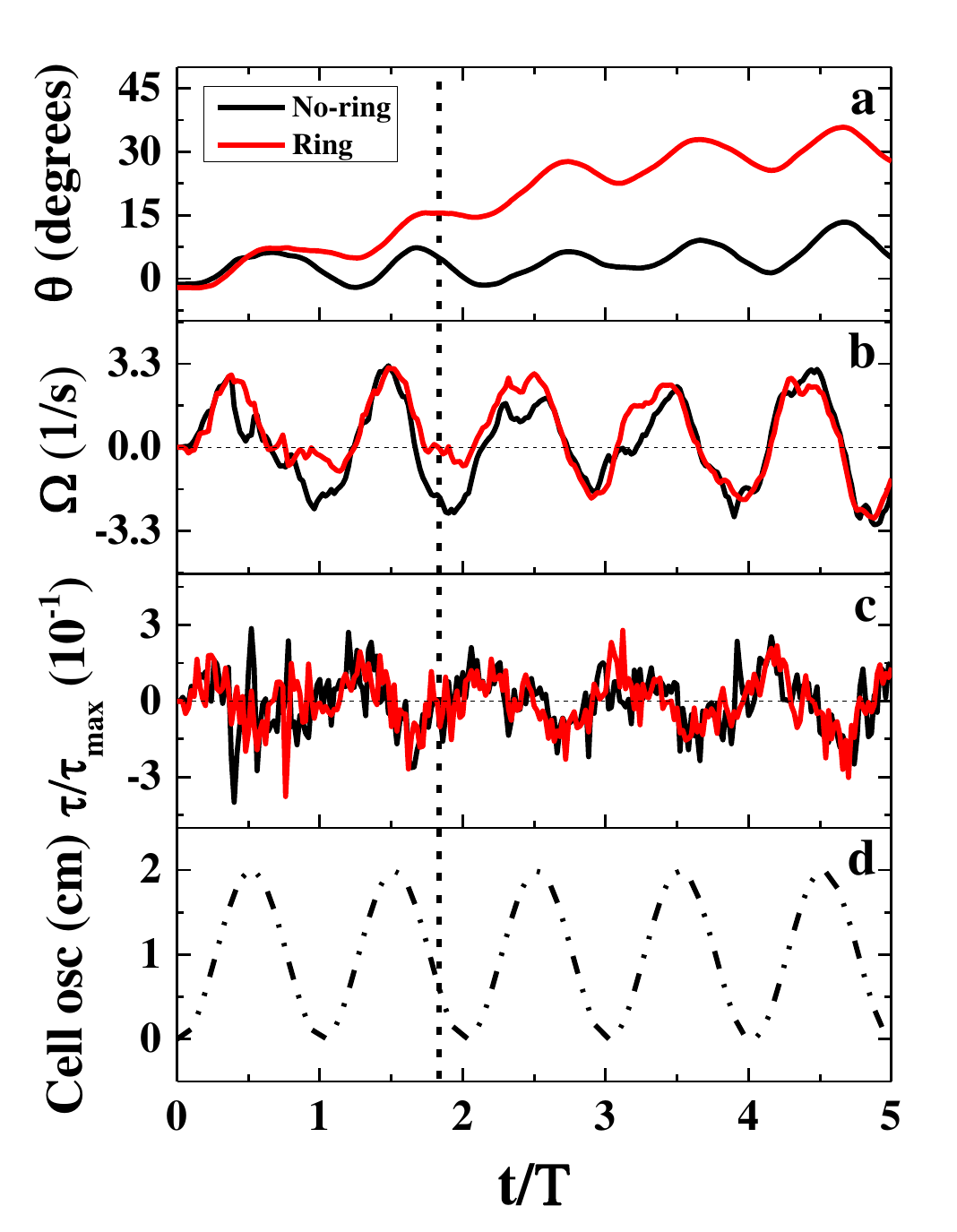}
\centering \caption{\label{Fig10}(color online) Simulations. Temporal evolution of the
calculated magnitudes for the intruders during the first second of a simulation 
using $ \Gamma = $ 1.0. From top to bottom: the evolution of the rotation $\theta$ (a),
the angular velocity $\Omega$ (b), the torque about the center of mass $\tau$ (c) and finally,
the position of the cell (d) showing the oscillation described by it. The vertical line indicates the instant $t =$ 1.84 T. $\tau_{max}$ is maximum torque on a non-tilting but horizontally accelerated No-ring intruder.}
\end{figure}

According to our simulations, the difference in tilting between intruders 
lies in that one type of intruder, the No-ring one, is somehow more capable of 
rectifying its rotation during sinking, while the other, the Ring intruder, 
is not. This rectification can be understood as the 
process of returning to, or recovering, the initial rotation angle once one 
oscillation of the cell has concluded and, as can be seen in the temporal 
evolution of $\theta$ (Fig. \ref{Fig10}(a)), the difference in the rotation
angles between the No-ring and Ring intruders is produced by a non-rectifying 
cumulative process taken by the latter.

To understand why the tilting dynamics is affected by the presence of the 
legs, which makes the No-ring intruder able to further rectify its rotation 
-- at least for the values of gamma between 0.25 and 1.25--, we calculate 
from the simulations the torque about the center of mass and the angular velocity.

Fig. \ref{Fig10} shows the time evolution of $\theta$, the 
angular velocity, the torque about the center of mass and the oscillation of 
the cell for the two types of intruders in a simulation with 
$\Gamma$ = 1.0. Focusing on the $\theta$ curve, the difference in 
tilting can be noticed during the first second of the simulation (as in 
Fig. \ref{Fig7} (b) for the averaged values) as well as the aforementioned 
Ring intruder non-rectification process. Unexpectedly, 
the values of the torques shown in Fig. \ref{Fig10} (c) are very similar 
for the two intruders, contradicting the intuitive idea that the sole 
presence of the legs would produce higher torques about the Ring intruder's center 
of mass making it rotate more. However, the slight 
differences may affect, to a greater or lesser extent, the rotational 
movement of the latter with respect to the No-ring intruder (see Fig. \ref{Fig10} (b)).

In general, during the first half of a cell oscillation, both intruders 
rotate in the same direction --counterclockwise due to the fact that the 
granular bed moves from left to right--. But, during the second half, 
some forces appear in the 
bottom and/or the inside of the Ring intruder's right leg that do 
not allow it to rotate in the same way as the No-ring does. In some 
cases, these forces completely prevent it from rotating clockwise, 
as in both first 
oscillations shown in Fig. \ref{Fig10}. This process is prone to occur 
in each of the oscillations during the simulation and its repetition causes 
the differences in inclination observed for the two intruders after 8 
simulated seconds (see Fig. \ref{Fig9}).

\begin{figure}
\includegraphics[width=0.45\textwidth]{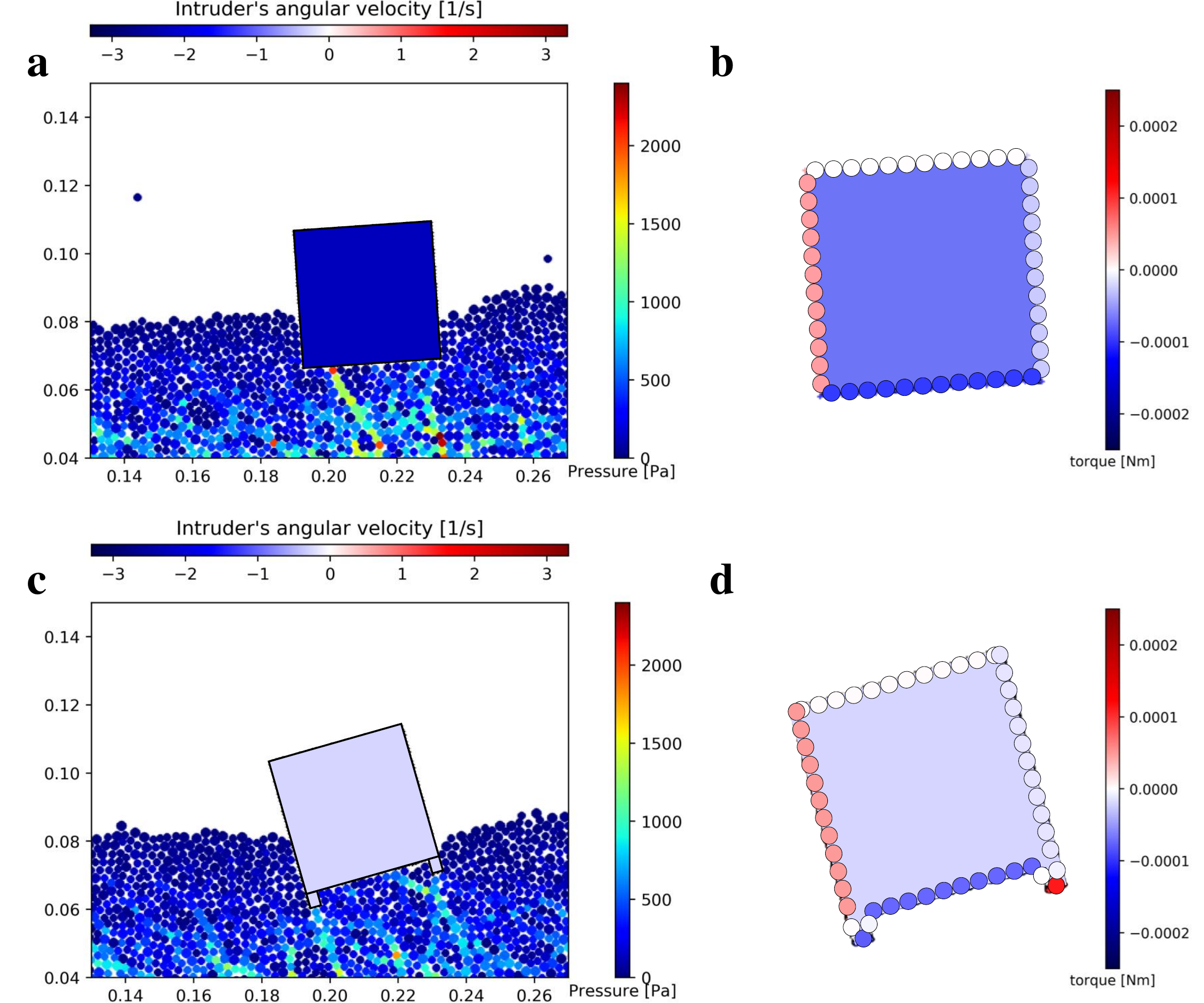}
\centering \caption{\label{Fig11}(color online) Simulations. Pressure field in the granular medium 
represented as a jet color map for No-ring (a) and Ring (c) intruders. 
The color of the intruders represents their angular velocity using a 
seismic color map. In (b) and (d) the outer segments of the intruders are represented as a whole with a color corresponding to the resulting torque about the center of mass that is generated on all grains belonging to them. Notice that for the Ring intruder each leg is divided into 3 segments. The grains in these segments are magnified for better viewing.  
The resulting torque about the center of mass is represented as the interior color of the intruder.
For both angular velocity and torque, blue color represents clockwise.}
\end{figure}

Fig. \ref{Fig11} illustrates in more detail what is described above. In (a) 
and (c) it shows the pressure field in the granular medium at $ t = $ 1.84 T (time indicated in Fig. \ref{Fig10} by the vertical line),
where the force chains are represented using a sequential color map. In them, 
the color of the intruders represents their angular velocity, which in turn 
is displayed as a diverging color map where blue indicates clockwise rotations. 
Figures \ref{Fig11} (b) and (d) show the contribution to the torque about the 
center of mass of each of the intruder segments as a result of the forces acting 
on them. In this case, the color of the intruders is associated with the resulting torque about the center of mass. Note that all the grains of each 
outer edge in (b) and (d) are represented with one color corresponding 
to the resulting torque about the center of mass obtained from the torques of all 
the particles on this same edge, though only the parts of the edge in contact with 
the granular medium are effectively interacting. These figures help identify what type of 
torque, clockwise or counterclockwise, is generated in each part of the 
intruders (including the legs), how it is generated, and how representative 
it is for the resulting torque about the center of mass. In this particular 
case, it is observed for the Ring intruder that the force associated 
with the lower part of the right leg is responsible for almost canceling the 
torque about the center of mass. Therefore, the Ring intruder remains inclined
while the No-ring one rotates back to its original position. 

\begin{figure}
\includegraphics[width=0.45\textwidth]{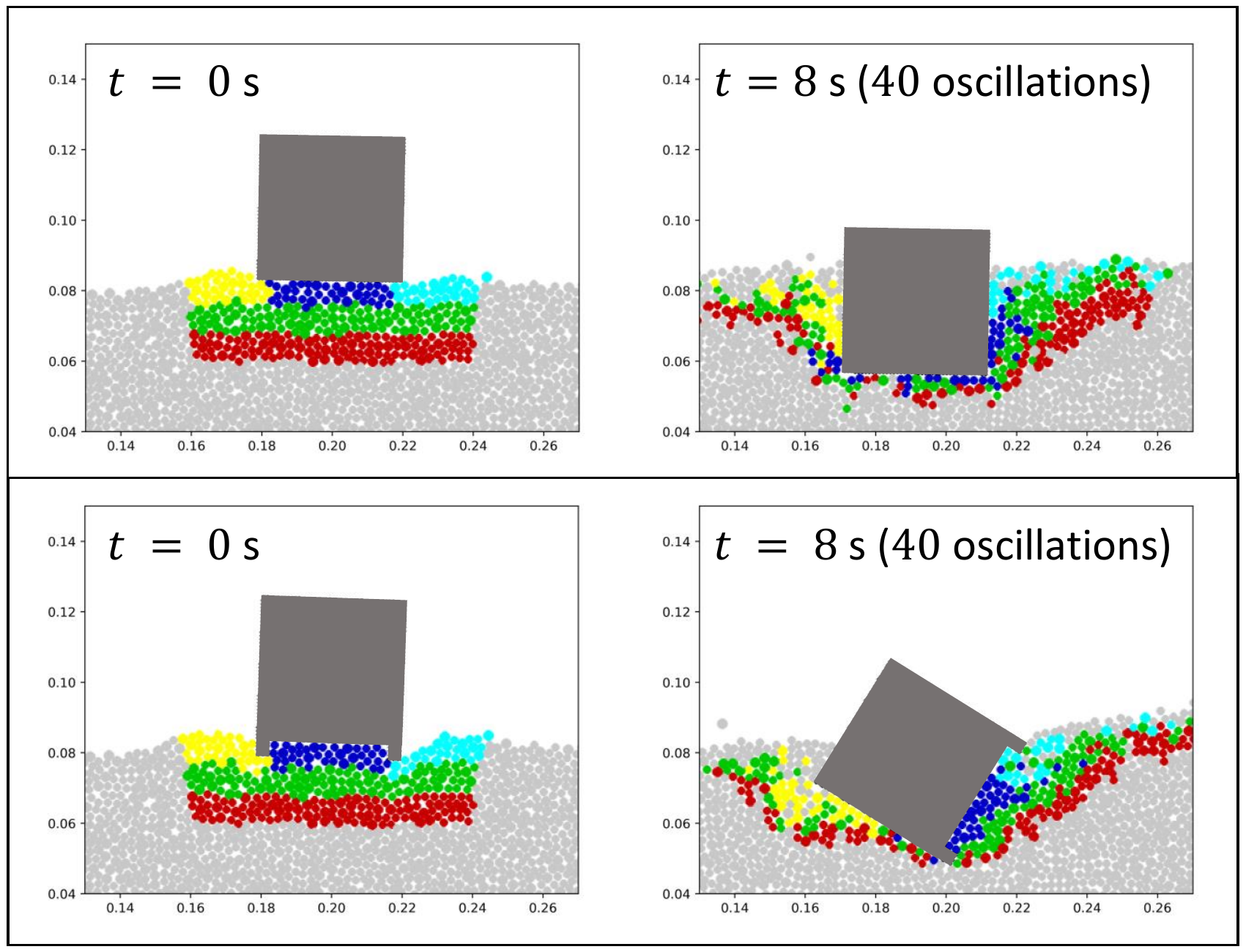}
\centering \caption{\label{Fig12}(color online) Simulations. Grain tracking. 
Initial (left) 
and final (right) positions of the No-ring (top) and Ring (bottom) intruders as 
well as the grains of the granular medium next to them. 
The grains right below the intruders are represented in blue while the yellow 
and cyan colors represent the grains initially at the left and 
right side of the intruders, respectively.
Note that for the Ring intruder, the blue grains initially located between the legs remain in the same region until the end of the simulation}.
\end{figure}

Those forces acting on the Ring intruder's legs could be associated with 
the grain jamming between them. Fig \ref{Fig12} shows that almost all 
the blue grains that were initially in the region between the legs remain there 
throughout the simulation. Furthermore, it can be seen 
that some of the grains (cyan) around the Ring intruder accumulate between its legs:
this is because during the first part of the oscillations these grains tend to 
move towards the legs, however, they cannot leave them during the second part. 
In contrast, grains in the 
region below No-ring intruders have more freedom to exit.
The previous process suggests that the Ring intruder along with the grains 
between the legs could be treated as a No-ring intruder with 
increased friction at the bottom. This increase in friction will make the 
grains underneath more likely to get stuck, preventing them from coming out 
and thus creating force chains capable of stopping the 
clockwise rotation (restoring towards vertical position) of the intruder.

A final observation from the simulations for
dimensionless accelerations of $ \Gamma = $ 1.5 is that the No-ring intruder 
rotates almost 90 degrees in a 100\% of the cases where it reached 45 degrees, 
doing so in an abrupt way. The Ring intruders, however, show a constant growth up 
and, in some cases (after reaching 90 degrees), increase the slope reaching values of up to 180 
degrees. It is worth noting that once Ring intruders turn 90 degrees they 
begin to resemble No-ring ones as the presence of the legs loses importance 
in the penetration dynamics. Behaviors such as those described before were not observed 
experimentally since the $ \Gamma $ values used in the experiments did not 
exceed 1.24 due to technical limitations of the shaker used.

The experimental findings are explained not only by the numerical simulations, 
but also by a Newtonian model developed in the Appendix \ref{App.1}.
This model is based in the force balance on a cylinder sinking in a granular medium. 
The forces considered in the 1D model are gravity, a frictional force proportional 
to velocity and a pressure like force, proportional to the depth $h$, as expressed 
in Eq. (A.13). 
Though this model does not include the degree of freedom associated with tilting, 
the consideration of its influence in the lineal and surface dimensions of the 
intruder is enough to explain why a tilted intruder reaches a final depth smaller 
than that reached by a non tilted intruder for a given $ \Gamma $.

\begin{figure*}
\includegraphics[width=0.75\textwidth]{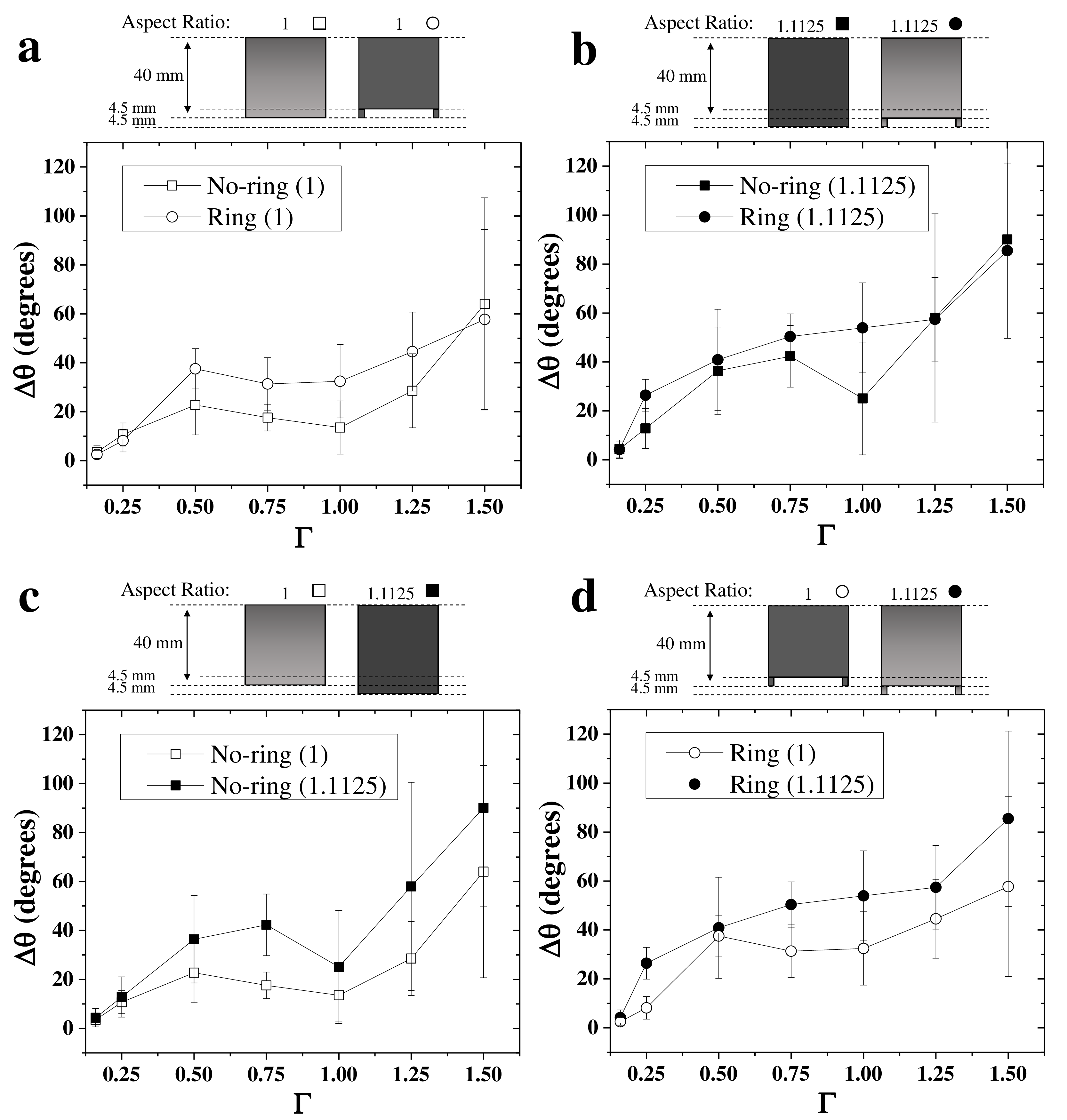}
\centering \caption{\label{Fig-123} Simulations. Final tilt angles at different $\Gamma$ for two intruders with (a) aspect ratio 1 and different foundations; (b) aspect ratio 1.1125 and different foundations; (c) a flat bottom and different aspect ratios and (d) a ring at the bottom and different aspect ratios.}
\end{figure*}

Now, we examine the influence of the intruder's aspect ratio on the penetration dynamics. The results of the simulations performed with the intruders of different aspect ratios are summarized in Fig. {\ref{Fig-123}}. Fig. {\ref{Fig-123}}(a) compares the dependence of the final tilt of two intruders with the same aspect ratio (1), one with legs and the other with a flat bottom. With the exception of the smaller values of $\Gamma$ where no noticeable differences are apparent, the intruder with legs always tilts more. This behavior is repeated in the results of Fig. {\ref{Fig-123}(b)} for a larger aspect ratio (1.1125): again the intruder with legs has a final tilt larger  than the one with a flat bottom. Therefore, the above suggests that the presence of a ring at the bottom of the intruder causes a higher final tilt, although it might become less important with increasing aspect ratio.
Figures {\ref{Fig-123}}(c) and (d) compare intruders with the same foundation and different aspect ratios. Both figures lead to the same conclusion: the larger the aspect ratio, the larger the tilt angle. 
Summarizing the results of Fig. {\ref{Fig-123}}, the intruder with higher aspect ratio and ring placed on the bottom has the largest tilt angle for all dimensionless accelerations, and the intruder with flat bottom and lower aspect ratio has the smallest tilt angle. Interestingly, the other two intruders show approximately equal final tilt angle values for equal values of $\Gamma$.

\section{Conclusions}
We have studied the behavior of cylindrical objects
as they sink into a dry granular bed fluidized by horizontal
oscillations, as a model system to understand the effects of
earthquake-related fluidization of soils on human constructions and
other objects like rocks.

We have found that, within a relatively large range of lateral
shaking amplitudes at a frequency of 5 Hz, cylinders with flat
bottoms sink vertically, while those with a \textquotedblleft
foundation\textquotedblright consisting in a shallow Ring attached
to their bottom, tilt laterally besides their vertical sinking.

We have been able to mimic the above described behaviors by quasi-2D
numerical simulations. With their help we found that these differences 
are not necessarily due to the sole existence of the legs that generate 
bigger torques about the center of mass. Instead, they can be 
associated with the jamming of the particles in the region between the 
legs, which can increase the friction at the bottom of the Ring intruder 
generating force chains capable of preventing the 
total recovery of its initial angle of rotation. 
Numerical experiments also helped to clarify the 
influence of the intruder aspect ratio on the tilt dynamics: of two 
intruders with the same foundation, the one with higher aspect ratio will 
have a larger tilt angle. So, the aspect ratio and the foundation type cooperate to establish the penetration dynamics of the intruder.

We have also reproduced the vertical sink
dynamics of cylinders with a flat base using a Newtonian equation of
motion for an object penetrating a fluidized layer of granular
matter, where the granular effective density increases with depth,
eventually reaching a solid phase. The same model allows to
understand the sinking even in the present of tilting (Appendix \ref{App.1}).

Finally, it is worth noting that preliminary experimental data and
quasi-2D numerical simulations suggest that, when strong enough
lateral shaking is applied, the tilting scenario tends to dominate
regardless the nature of the intruder's foundation.

\begin{acknowledgments}
L. A. thanks the Photovoltaic Research Lab of the University of Havana 
for allowing the use of its facilities. E. A. drew inspiration from the late M. \'Alvarez-Ponte. We acknowledge support from Project 29942WL (Fonds de Solidarit\'e Prioritaire France-Cuba), from the EU ITN FlowTrans, and from the INSU.  We thank the SCAC of the French Embassy in Havana and the University of Strasbourg for their support. RT acknowledges the support of the Research Council of Norway through its Centres of Excellence funding scheme, Project No. 262644. We thank Mustapha Meghraoui, Einat Aharonov, Knut Jørgen Måløy, Eirik G. Flekkøy, for fruitful discussions. This research was made in the frame of the University of Havana's institutional project "Medios granulares: creando herramientas para prevenir catástrofes".
\end{acknowledgments}

\appendix*
\section{}
\subsection{Sink dynamics: a phenomenological Newtonian model}
\label{App.1}

The model to be formulated should account for two related processes,
the sinking in the  vertical direction and the oscillations perpendicular 
to it. But as was shown above, the No-ring intruders have only small 
oscillations that end fast, being the overall sinking process almost 
vertical. Regarding the Ring intruders, though they strongly oscillate,
the tilting  process ends first, so we will consider only the equation 
controlling the vertical sinking, figuring out how the tilting angle affects 
the sinking dynamics.

In order to formulate a model to describe analytically the sinking process,
let us consider the forces acting on the cylinder. As soon as the shaking
starts, if the dimensionless acceleration is above threshold, the upper
part of the granular bed is fluidized, and the intruder sinks.

Let us assume that the cylinder just sinks vertically, and let us
name the vertical downward axis as $z$. The force balance on the
intruder can be written as

\begin{equation}
 \label{eq.2}
 m \vec a = m \vec g + \int (-P)  \hat{n} dS +\int  \sigma_s \cdot \hat{n} dS
\end{equation}

\noindent where $P$ is the pressure, $\sigma_s$ the shear stress
tensor, $\hat{n}$ is the vector normal to the intruder's surface,
and the integrals run over the boundary of the intruder that is
inside the granular material. Assuming a hydrostatic pressure
profile, we can write:

\begin{equation}
 \label{eq.3}
  P= \int_0^h \rho(z') g dz'
\end{equation}

\noindent where $h$, as previously, is the depth reached by the
cylinder below the surface of the granular medium. In
Eq. (\ref{eq.3}) we have made explicit that the density of the
material varies with depth. Let us assume that it varies as a power
law between zero and the density of the solid layer, $\rho_{sl}$,
that is reached at a depth $h_f$:

\begin{equation}
 \label{eq.4}
 \rho(z')=\rho_{sl} \Big ( \dfrac {z'} {h_f} \Big )^p
\end{equation}

\noindent where $p \in [0,1]$. The selection of the value of $p$ is discussed 
below (see also Appendix \ref{App.2}).

By combining (\ref{eq.4}) and (\ref{eq.3}) and integrating, we find
the hydrostatic buoyancy force acting on the cylinder with a length
$h$ under the (average) level of the granular bed, as:

\begin{equation}
 \label{eq.5}
 \int (-  {P})  \hat{n} dS=- \dfrac {\rho_{sl} S g } {(p+1) h_f^p} h^{p+1} \hat h
\end{equation}

\noindent where $S$ is the characteristic area of the intruder
cross section, and $\hat h$ is a unit vector pointing downwards. It is
easy to see that the buoyancy force depends on the volume submerged
into the granular medium.

Neglecting the inertial forces, which according to our simulations is 
typically two orders of magnitude smaller than the contact forces, the 
shear stress component goes as

\begin{equation}
 \label{eq.6}
\int \sigma_s \cdot \hat n dS = -D \gamma v \hat h
\end{equation}

\noindent where $\gamma$  has the dimensions of a viscosity, $D$ is
the characteristic size of the cross section of the intruder and $v$ is
its sinking speed \cite{deBruyn2004, Hou2005}. By substituting Eq. 
(\ref{eq.5}) and Eq. (\ref{eq.6}) into Eq. (\ref{eq.2}), and only 
recovering the modular values, we get:

\begin{equation}
 \label{eq.7}
m \dfrac {d^2h} {dt^2}+ D \gamma \dfrac {dh} {dt} + \dfrac {\rho_{cs} S g } {(p+1) h_f^p} h^{p+1} =mg
\end{equation}

\begin{figure}
\includegraphics[width=0.4\textwidth]{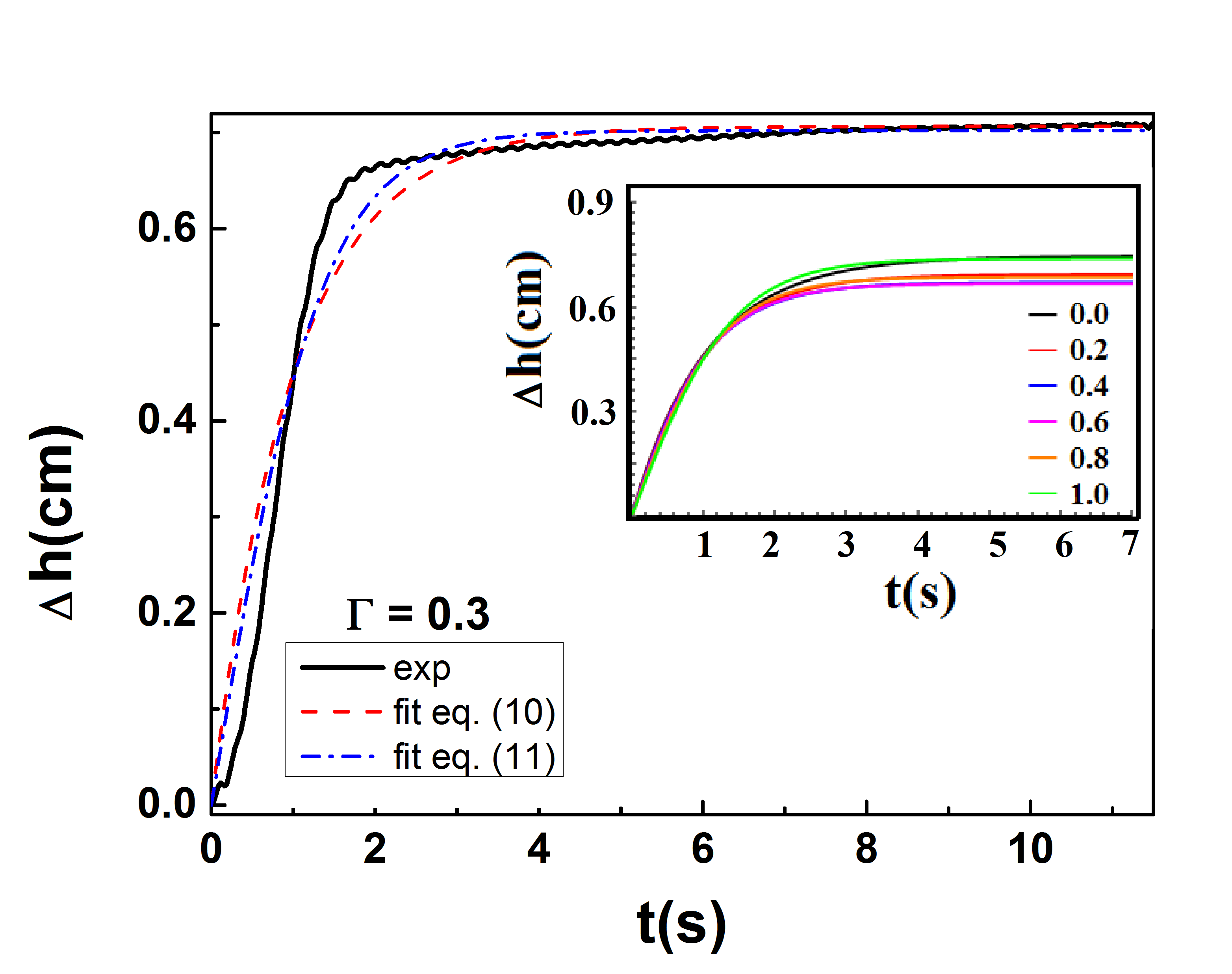}
\centering \caption{\label{Fig-A1} (color online) Time dependence of sinking depth
for the No-ring cylinder from experiment, compared with that
determined from Eqs. (\ref{eq.10}, \ref{eq.11}). The inset shows the
solutions of Eq. (\ref{eq.7}) for different values of $p$ (see
text).} 
\end{figure}

Before solving Eq. (\ref{eq.7}) we will assume that the sink
velocity is constant, which follows quite well the behavior during
the fast sink regime, as seen in Fig. \ref{Fig-A1} ({\it i.e.}, we
neglect the inertial term). So,

\begin{equation}
 \label{eq.8}
 \dfrac {dh} {dt} + \dfrac {\rho_{sl} S g } {D \gamma (p+1) h_f^p} h^{p+1} =\dfrac {mg} {D \gamma}
\end{equation}

\noindent which can be written as

\begin{equation}
 \label{eq.9}
 \dfrac {dh} {dt} + a h^{p+1} =b
\end{equation}

The definitions of $a$ and $b$ are easily deduced by comparing Eqs. (\ref{eq.8}) and
(\ref {eq.9}).

Equation (\ref{eq.9}) has analytical solutions if $p=0$ or $p=1$,
which correspond to the extreme cases of constant density and a
linear density profile with depth, respectively. The solutions are

 \begin{equation}
\label{eq.10}
h(t)=\dfrac b a (1-e^{-a t})
\end{equation}

\noindent if $p=0$, and

 \begin{equation}
\label{eq.11}
h(t)=\sqrt {\dfrac b a} \tanh (\sqrt {a b} t)
\end{equation}

\noindent if $p=1$.

It is easy to see that both expressions correspond to an exponential
growth that saturates.

Figure \ref{Fig-A1} shows the experimental results (continuous line)
obtained for a dimensionless acceleration $\Gamma \simeq 0.3 $. It
is possible to see in more detail the initial fast sinking process,
followed by the slow creep.  Fig. \ref{Fig-A1} also shows the fitting
of equations (\ref{eq.10}, \ref{eq.11}) to experimental data. Both 
solutions reproduces well the main features of the sinking process.

It is almost impossible to determine experimentally the exact
density profile. But we do not need to know it in order to validate
our model, if we use the following rationale. Firstly, we fit Eqs.
(\ref{eq.10}, \ref{eq.11}) to the experimental data and obtain the
values of $a$, $b$ that correspond to $p=0$ ($a(p=0),b(p=0)$) and
$p=1$ ($a(p=1),b(p=1)$). Let us assume that $a$ and $b$ vary
linearly with $p$ between the extremes values which were obtained
from the fitting process. For an intermediate value of $p$ (say,
$p_1$) we can calculate the corresponding values of $a(p_1)$ and
$b(p_1)$. With them, we can in turn determine the constants of Eq.
(\ref{eq.7}). Then, we solve this equation numerically. This
procedure is repeated for values of $p$ between 0 and 1, with a step
of 0.1.

The inset in Fig. \ref{Fig-A1} shows some of the numerical solutions
for the values of $p$ in the legend. The main conclusion is that the
density profile has small influence on the first (and most
important) part of the sinking process. Of course, the final depth
is influenced by the value of $p$, but due to experimental
uncertainties, it is almost impossible to choose any particular
value.

Let us now study the influence of the values of $p$ in the quality
of the fit of the solution of Eq. (\ref{eq.7}) to the experimental
data. For doing this we notice that the values of $a$ and $b$ in Eq.
(\ref{eq.10}) can be easily obtained from the experiments.
Considering Eq. (\ref{eq.9}) in the first moments of motion, as $h$
is small, $h'(t) \simeq b$, so $b$ can be evaluated as the initial
slope. As at large times $h(t)\sim h_{eq}$ ($h_{eq}=\Delta h $ if
$h(0)=0$) then $a=b/h_{eq}^{p+1}$. Then solving Eq. (\ref{eq.7}) for a
given value of $b, p, a(p)$ and naming the result $h_{mod}$, the
best value of $p$ arises from the minimization:

\begin{equation}
\label{eq.12}
p_{opt}=\arg\min \sum_{i=1}^N [h_{mod}(t_i,p)-h_{exp}(t_i)]^2
\end{equation}

\noindent where $h_{exp}(t)$ are the experimental values of $h$.

The result for $\Gamma \leq 1.0$ is indifferent to $p$: the fit is
equally good no matter which is the value of $p \in [0,1]$. For
$\Gamma = 1.24$ there are differences in the quality of the fits for
various values of $p$, but Eq. (\ref{eq.12}) gives a minimum for $p=0$,
so, we will use $p=0$ in the following (Appendix \ref{App.2} supports 
the selection of $p$ from the simulations). Then, Eq. (\ref{eq.7})
becomes:

\begin{equation}
 \label{eq.13}
m \dfrac {d^2h} {dt^2}+ D \gamma \dfrac {dh} {dt} + \dfrac {\rho_{sl} S g } { h_f} h =mg
\end{equation}

\noindent that can be taken as the simplest equation of motion describing
the vertical sink dynamics of our cylinders. It is worth
noticing that Eq. (\ref{eq.13}) reproduces quite closely the results
reported in Fig. \ref{Fig-A1}, and can be used to qualitatively describe
the vertical sinking of Ring-cylinders while tilting, as we will see below.
It is possible to demonstrate that Eq. (\ref{eq.13}), developed for
a granular bed fluidized by shaking, is closely related with that proposed in
\cite{Pacheco2011} to describe the penetration of an intruder into
ultra-light granular material that eventually behaves like a fluid
medium even in the absence of shaking.

In order to understand the influence of the tilting dynamics
in the sinking process, it is useful to note that, when applying Eq. 
(\ref{eq.13}) to a tilted cylinder, the values of both $D$ and $S$ 
change. The reason is that when we calculate the surface integrals, 
the result will be proportional to the cylinder's immersed volume. 
As the cylinder tilts, both the immersed surface and linear dimensions 
increase more than in the case of sinking without tilting,
so the drag force is bigger in the former case. Considering, for
instance, the situation represented in the lower row of  Fig.
\ref{Fig2}, when the cylinder sinks a distance  $\Delta h $, the 
surface and linear dimensions increase as the inverse of $\cos \theta$ 
(of course, other intruder geometries may follow different laws).

To test it, let us assume a simplified model: the increase factor of
$S$ and $D$ is proportional to the characteristic size of the cross
section of the cylinder projected on the horizontal plane, i.e., it
is proportional to the inverse of $\cos \theta$. Then, instead of
$D$ and $S$, we will solve Eq. (\ref{eq.13}) using $D / \cos \theta (t)$
and $S / \cos \theta(t)$, where $\theta (t)$ is a function that grows 
from zero to the maximum angle $\theta_{max}$ reached by the cylinder, 
mimicking Fig. \ref{Fig5}(c), i. e. with the same time constant.

\begin{figure}
\includegraphics[width=0.4\textwidth]{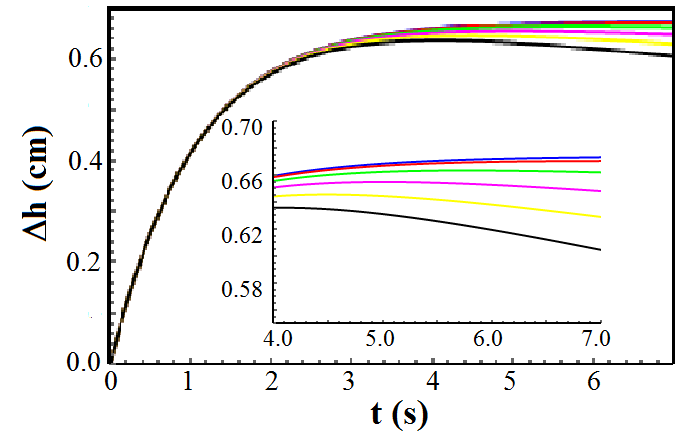}
\centering \caption{\label{Fig-A2} (color online) Time dependence of sinking depth
as calculated solving numerically Eq. (\ref{eq.13}) considering the variation of
$S$ and $D$ provoked by tilting (see text). Upper
curve is for $\theta_{max}=$ 0 while the lower one is for $\theta_{max}= \pi /
3$. Between them, $\theta_{max}$ varies in steps of $\pi/15$. The inset shows
the last three seconds.
}
\end{figure}

The consequences can be seen in Fig. \ref{Fig-A2}. While in the beginning 
the sinking process in all situations occurs with the same dynamics, as 
the cylinder approaches the final angle, the behavior changes, being the 
final depth larger for the situations corresponding to low  tilting.

The upper curve, calculated for $\theta=$ 0 coincides with the upper
curve in the inset of Fig. \ref{Fig-A1} (calculated for $p=0$). Subsequent curves are
calculated for values of $\theta_{max}$ varying in steps of $\pi /$15, the
lowermost curve corresponds to $\theta_{max}=\pi/$3. As the inclination of
the cylinder increases, both the buoyancy and the viscous drag do.
The effect of these factors on the sinking process of Ring cylinders
was already noted in Fig. \ref{Fig5}(b): an immediate consequence is
the decrease of the sinking depth (for a given $\Gamma$), compared with
that of the No-ring ones, which can be easily observed in
the experiments. From the inset it is possible to deduce that, for the
larger angles, a small decrease in the depth is observed.

In spite of the simplifications assumed, it is worth noting that one of the 
basic differences between Fig. \ref{Fig6} (a) and (b) --for a given $\Gamma$ the 
No-ring cylinders sink deeper than the Ring ones-- could be qualitatively 
described by our model.

Finally, there is another element that was neither considered by us:
as the container shakes horizontally, it produces a
horizontal drag that changes periodically its direction. According
to \cite{Zhang2015}, it creates an additional lift force, and also a
dependence of the drag force with depth, which, of course, must
influence the detailed penetration dynamics of the Ring cylinders. The results 
of Li et al \cite{Li2013} also support these ideas.

\subsection{Density profile of the granular medium}
\label{App.2}

At each time step the simulation box is subdivided into a fixed number of 
rectangles for which the density is calculated as $\rho = m / V$, where $m$ is the sum 
of the masses of all the particles of the granular medium within each rectangle and $V$ is the volume of 
the rectangle calculated as its area multiplied by the average diameter 
of the particles of the medium inside it.

\begin{figure}
\includegraphics[width=0.4\textwidth]{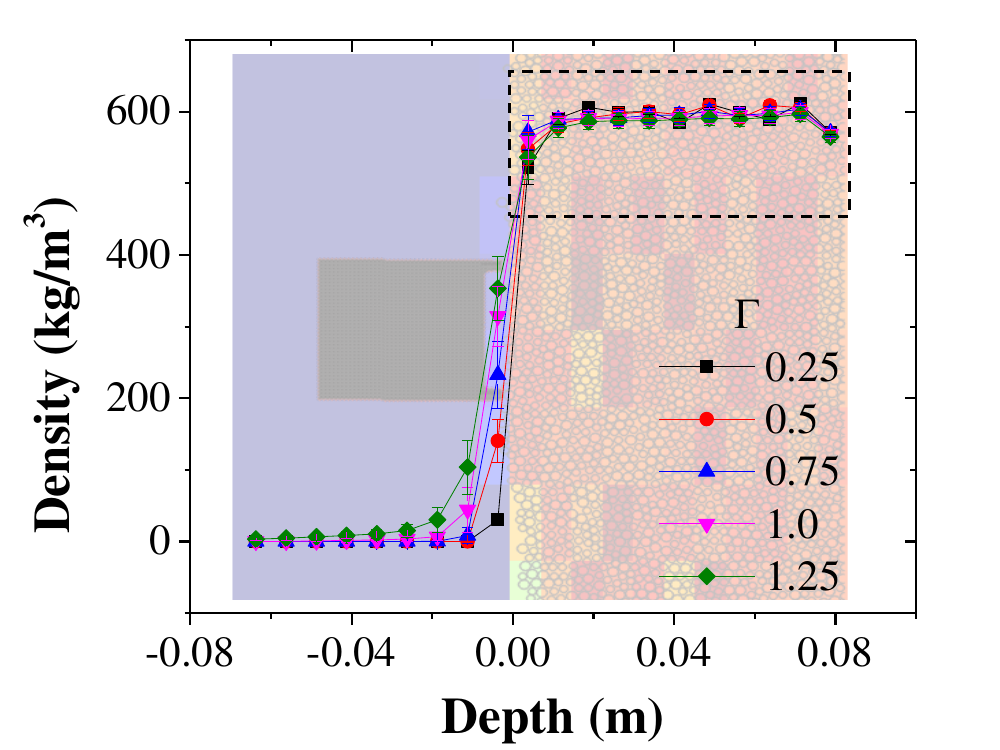}
\centering \caption{\label{Fig-A2.2} (color online) Density profiles for the different values
of dimensionless acceleration $\Gamma$ obtained from the simulations. The background shows 
the subdivision of the simulation box where the color of each rectangle goes sequentially
from blue to red and represents its density. The intruder is represented in light gray.
}
\end{figure}

The density profile for each dimensionless acceleration $\Gamma$ is obtained by calculating the average 
of the density profile at each time step. The latter is obtained by averaging 
the profiles of a set of rectangle columns near the intruder. 
These columns are chosen avoiding the presence of the intruder in them in order not 
to affect the density profile since only the particles that compose the granular 
medium are taken into account in its calculation.

Figure \ref{Fig-A2.2} shows the density profiles as a function of depth for the
different values of $\Gamma$. As can be seen in the region of the graph enclosed 
by dashed lines, the density of the medium saturates rapidly with increasing depth 
for all dimensionless accelerations. This fact supports from the simulations the use of $p = 0$ in expression \ref{eq.4} of Appendix \ref{App.1}.

\bibliography{Phys-Rev-E}

\begin{thebibliography}{60}
\bibitem [{lim()}]{limit}%
  \BibitemOpen
  \href@noop {} {}\bibinfo {note} {The maximum oscillation amplitude that the
  shaker can generate is 13 mm. For a frequency of 5.0 Hz, the maximum possible
  acceleration is 12.83 m/s$^2$}\BibitemShut {NoStop}%
\end{thebibliography}

\end{document}